\documentclass[aps,prx,twocolumn,superscriptaddress,floatfix,10pt]{revtex4-2}
\usepackage{enumitem}
\usepackage[utf8]{inputenc}
\usepackage[english]{babel}
\usepackage[T1]{fontenc}
\usepackage{amsmath}
\usepackage{feynmf}
\usepackage{physics}
\usepackage[table]{xcolor}
\usepackage{soul}
\usepackage{graphicx}
\usepackage{amsfonts}
\usepackage{blindtext}
\usepackage{appendix}
\usepackage[colorlinks = true,
            linkcolor = blue,
            urlcolor  = blue,
            citecolor = blue,
            anchorcolor = blue]{hyperref}
\usepackage{siunitx}
\usepackage{adjustbox}

\usepackage{nicefrac}

\usepackage{booktabs}
\usepackage{qcircuit}
\usepackage{soul}
\usepackage{rotating}

\usepackage{hyperref}
\usepackage{orcidlink}

\usepackage{algorithm}
\usepackage{algpseudocode}
\usepackage{amsmath}

\usepackage{stmaryrd}

\usepackage{tabularx}
\usepackage[capitalise,noabbrev]{cleveref}

\usepackage{algorithm}
\usepackage{algpseudocode}
\algnewcommand\Input{\item[\textbf{Input:}]}
\algnewcommand\Output{\item[\textbf{Output:}]}

\newcommand{\CZ}{{\sf{CZ}}\xspace}

\newcommand{\CX}{{\sf{CX}}\xspace}
\newcommand{\CY}{{\sf{CY}}\xspace}
\newcommand{\LC}{{\sf{LC}}\xspace}

\newcommand{\nkd}[3]{\left \llbracket #1,\, #2,\, #3 \right \rrbracket}

\usepackage[normalem]{ulem}

\begin{abstract}
    We propose a general method for preparing stabilizer states with reduced two-qubit gate count and depth compared to the state of the art. 
    The method starts from a graph state representation of the stabilizer state and iteratively reduces the number of edges in the graph using two-qubit Clifford gates to produce a unitary preparation circuit. 
    We explore various heuristic search and AI-based approaches to optimally choose Clifford gates at each step, the most sophisticated of which is a combination of reinforcement learning and Monte Carlo tree search that we call \textit{QuSynth}. We apply our method to synthesize code states of various quantum error correcting codes including the 23-qubit Golay code and the 144-qubit gross code, the latter of which is significantly beyond {the qubit number that is accessible to} prior optimal circuit synthesis methods. { We demonstrate that our techniques are capable of reducing the required} two-qubit gates by up to a factor of 2.5 compared to previous approaches while retaining low circuit depth.  
\end{abstract}

\begin{document}
\title{Fast stabilizer state preparation via AI-optimized graph decimation}

\author{Michael Doherty\,\orcidlink{0009-0009-2904-4553} $^{\dagger}$}
\thanks{These authors contributed equally to this work. \\ $^{\dagger}$ Work carried out during an internship at Quantinuum, Partnership House, Carlisle Place, London SW1P 1BX, United Kingdom.}
\affiliation{University College London, Torrington Place, London WC1E 7JE, United Kingdom}

\author{Matteo Puviani\,\orcidlink{0000-0002-5332-213X}}
\thanks{These authors contributed equally to this work. \\ $^{\dagger}$ Work carried out during an internship at Quantinuum, Partnership House, Carlisle Place, London SW1P 1BX, United Kingdom.}
\affiliation{Quantinuum, Partnership House, Carlisle Place, London SW1P 1BX, United Kingdom}

\author{Jasmine Brewer\,\orcidlink{0000-0002-3084-0663}}
\thanks{These authors contributed equally to this work. \\ $^{\dagger}$ Work carried out during an internship at Quantinuum, Partnership House, Carlisle Place, London SW1P 1BX, United Kingdom.}

\author{Gabriel Matos\orcidlink{0000-0002-3373-0128}}
\affiliation{Quantinuum, Partnership House, Carlisle Place, London SW1P 1BX, United Kingdom}

\author{David Amaro\orcidlink{0000-0001-7853-9581}}
\affiliation{Quantinuum, Partnership House, Carlisle Place, London SW1P 1BX, United Kingdom}

\author{Ben Criger\,\orcidlink{0000-0001-9959-6462}}
\affiliation{Quantinuum, Terrington House, 13-15 Hills Rd, Cambridge, CB2 1NL, UK}
\affiliation{Institute for Globally Distributed Open Research and Education (IGDORE)}

\author{David T. Stephen\,\orcidlink{0000-0002-3150-0169}}
\affiliation{Quantinuum, 303 S. Technology Ct., Broomfield, Colorado 80021, USA}

\maketitle

\section{Introduction}
\label{sec:introduction}

Quantum state preparation is fundamental to quantum computing, serving as the foundation for quantum algorithms, error correction protocols, and simulation. 
The efficiency of state preparation circuits, typically measured by two-qubit gate count and circuit depth, directly impacts the performance of near-term and fault-tolerant quantum applications, where gate errors and decoherence impose resource constraints.
In this work, we focus on high-efficiency preparation of stabilizer states, a class of quantum states used in a variety of quantum algorithms \cite{Hein2006}, quantum networking \cite{Englbrecht2022transformationsof}, measurement-based quantum computation \cite{Raussendorf2003measurementbased}, and as logical states in quantum error correction \cite{gottesman1997stabilizer}.

Several recent works have explored optimization techniques for circuit synthesis problems, including heuristic search for CNOT and Clifford circuit synthesis \cite{webster2025}.
Reinforcement learning (RL) in particular has recently been used for problems ranging from discovery of QEC codes and encoding circuits \cite{Olle2024}, to fault-tolerant state preparation \cite{Zen2025}, Clifford synthesis \cite{kremer2025practicalefficientquantumcircuit,11032620} optimization \cite{foesel2021, rosenhahn2025}, and Pauli network synthesis \cite{dubal2025paulinetworkcircuitsynthesis}.
Although these recent works treat problems similar to stabilizer state preparation, they are restricted to systems of at most $\sim 20$ qubits.
Previous work on graph state preparation with RL \cite{giordano2025hybridrewarddrivenreinforcementlearning} has been limited to 7-qubit codes.
Algebraic methods \cite{bataille2021reducedquantumcircuitsstabilizer} are designed for dense graphs and perform poorly on the moderate-density graphs typical of QEC codes.

In this work, we simplify the state synthesis problem using the graph state representation of stabilizer states.
Every stabilizer state is \emph{locally Clifford-equivalent} to a graph state \cite{VanDenNest2004}, meaning that methods to prepare graph states can prepare arbitrary stabilizer states with the same two-qubit gate count and two-qubit depth.
The graph state representation allows us to transform the circuit synthesis problem into a sequential decision-making problem on graphs that we call \emph{graph decimation}. This enables the application of simple artificial intelligence algorithms for sequential decision making, including best-first search, beam search, and Monte Carlo tree search. 
We develop \textit{QuSynth}, a method that uses reinforcement learning in tandem with Monte Carlo tree search to prepare stabilizer states with low two-qubit gate count.
We demonstrate that our approach far outperforms the current state of the art in terms of two-qubit gate count and circuit depth, and scales up to large codes, including the $144$-qubit Gross code.
Our highly-optimized heuristic search methods also obtain solutions of similar quality to \textit{QuSynth} in a fraction of the time.

This paper is organized as follows: In \cref{sec:theory}, we introduce graph states, and the effects of elementary Clifford gates on these states.
Using these properties, we express graph state synthesis as a graph decimation problem, which is naturally solved with AI methods. \cref{sec:heuristic_method} develops search methods based on classical heuristics that serve as a benchmark for machine learning methods. We present our \textit{QuSynth} approach in \cref{sec:ml_methods} and benchmark it against traditional reinforcement learning. The main results of this paper are shown in \cref{table_results} and discussed in \cref{sec:results}.
We present possible future developments and applications, including possible routes to fault-tolerant state preparation, in \cref{sec:discussion} before concluding in \cref{sec:conclusion}.

\section{Graph States}
\label{sec:theory}

Our method is based on the representation of stabilizer states as graph states \cite{VanDenNest2004}.
First, recall that a general stabilizer state $\ket{\psi}$ on $N$ qubits is described by a set of $N$ independent commuting Pauli operators, called stabilizers, and denoted $S_i$ for $i=1,\dots,N$ such that $S_i\ket{\psi} = \ket{\psi}$ \cite{gottesman1997stabilizer}.
A graph state is defined by a graph $G$ consisting of nodes $i\in V$ and edges $\{i,j\} \in E\subset [V]^2$, which are identified by unordered pairs of nodes.
The corresponding graph state is
\begin{equation}
    \ket{G} = \left(\prod_{\{i,j\} \in E} \CZ_{i,j} \right) \bigotimes_{i\in V} \ket{+}_i,
\end{equation}
where $\CZ_{i,j} = \ketbra{0}_i\otimes I_j+ \ketbra{1}_i\otimes Z_j$ is the controlled-$Z$ gate. 
A graph state is a stabilizer state with stabilizers defined by
\begin{equation} \label{eq:graph_stabilizers}
    g_i = X_i \prod_{j\in n(i)} Z_j,
\end{equation}
such that $g_i\ket{G} = \ket{G}$ for each $i\in V$, where $n(i) = \{j \,\vert\, \{i,j\}\in E\}$ is the neighborhood of $i$.

In Ref.~\cite{VanDenNest2004}, it was shown that every stabilizer state is equivalent to a graph state, up to local Clifford gates.
That is, for any stabilizer state $\ket{\psi}$, there exists a graph $G$ and product of single-qubit Cliffords $C=\prod_{i\in V} C_i$, such that
\begin{equation}
\label{eq:stabilizer_to_graph}
    \ket{\psi} = C\ket{G}.
\end{equation}
Therefore, when constructing state preparation circuits for stabilizer states, it is sufficient to only consider graph states, as the local Clifford gates contained in $C$ can be easily applied after preparing the corresponding graph state.

\begin{figure}
    \centering
    \includegraphics[width=0.9\linewidth]{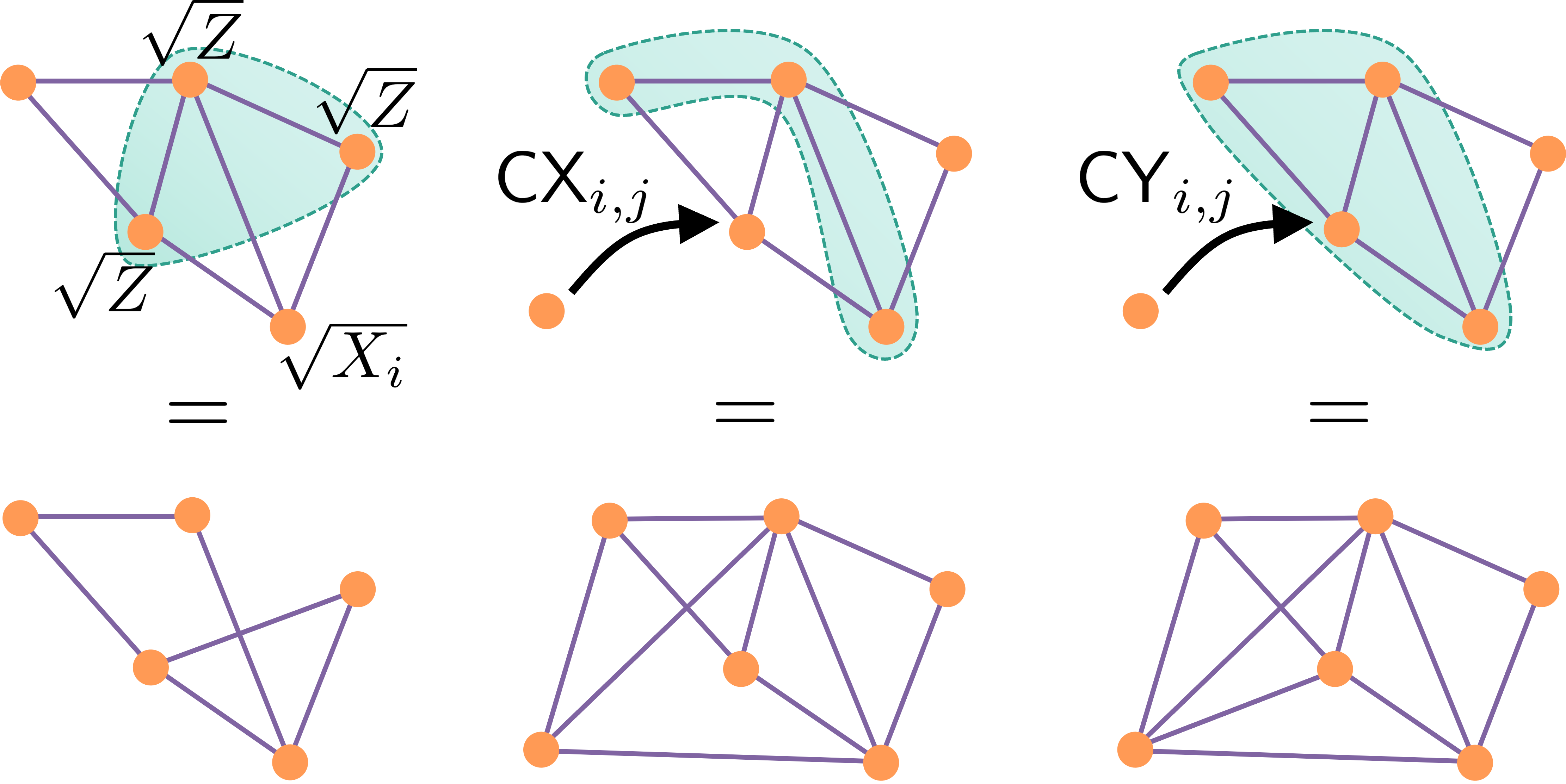}
    \caption{Transformations of graph states. Each circle represents a qubit initialized in $|+\rangle$, and solid edges denote $\CZ$ gates. Left: Illustration of the local complementation operation, where the highlighted region denotes $n(i)$. Middle: Illustration of \cref{eq:cx_derivation} where the arrow point from control qubit to target qubit and the highlighted region denotes $n(j)$. Right: Same as previous panel for \cref{eq:rule_y} where the highlighted region now denotes $n(j)\cup \{j\}$.}
    \label{fig:main_rules}
\end{figure}

The representation of stabilizer states as graph states is not unique, meaning that different pairs $G$ and $C$ can define the same state $\ket{\psi}$.
One way to produce an inequivalent graph representing the same state is by using the \textit{local complementation (LC)} operation.
Given a node $i$, we can apply the local Clifford operation
\begin{equation}
    \LC_i := \sqrt{X_i}\prod_{j\in n(i)} \sqrt{Z_j}
\end{equation}
which results in
\begin{equation} \label{eq:lc}
    \LC_i\ket{G} = \ket{\LC_i (G)},
\end{equation}
where the graph $\LC_i (G)$ is obtained from $G$ by complementing all edges between the nodes contained in $n(i)$, \textit{i.e.}, adding edges which are missing and removing them if they are already present in $G$, as shown in \cref{fig:main_rules}, left.
By applying a sequence of local complementations to change $G$ (updating $C$ accordingly), it is possible to explore the \emph{LC-orbit} of the stabilizer state, which is a large space of inequivalent graphs representing the same state, up to local Cliffords \cite{VanDenNest2004}. 

\subsection{Graph state preparation circuits}
\label{circuits}

Given a stabilizer state $\ket{\psi}$ corresponding to a graph $G$ such that $\ket{\psi} = C \ket{G}$ as in~\cref{eq:stabilizer_to_graph}, one can \emph{na\"{i}vely} obtain a circuit that prepares it as follows:
\begin{enumerate}
    \item Initialize a quantum circuit in the $\ket{0}$ state, with one qubit for each node in the graph.
    \item Apply Hadamard gates to all qubits.
    \item Apply a \CZ gate between every pair of qubits corresponding to nodes connected by an edge in the graph $G$.
    \item Apply the set of single-qubit Clifford gates $C$.
\end{enumerate}
In such a circuit, the number of two-qubit gates is equal to the number of edges in the graph, which we denote as $|E|$.
Since these gates commute, the two-qubit gate depth is at most $\delta + 1$, where $\delta = \mathrm{max}\{|n(i)|,i\in V\}$ is the maximum degree of the graph, which is equal to the maximum number of two-qubit gates acting on a single qubit \footnote{This follows from Vizing's theorem \cite{vizing1964estimate}, which says that a graph can be colored with at most $\delta+1$ colors. Each color defines a disjoint circuit layer.}.
One can then try to minimize the two-qubit gate count (and/or depth) by searching the set of all graphs that can be obtained by a sequence of local complementations, finding the graph with minimal edge count $|E|$ (and/or minimal degree $\delta$).
In practice, this space is so large that one typically requires metaheuristic methods, such as simulated annealing or genetic optimization \cite{krueger2022vanishing,sharma2025minimizingnumberedgeslcequivalent} to reduce edge count.
Several works have employed this approach to simplify graph state preparation \cite{hoyer2006resources,Cabello2011,hahn2019quantum,Duncan2020graphtheoretic,Adcock2020mappinggraphstate,bataille2021reducedquantumcircuitsstabilizer,kaldenbach2024mapping,vijayan2024compilation,sunami2022graphixoptimizingsimulatingmeasurementbased,Lee2023graphtheoretical,li2022photonic,sharma2025minimizingnumberedgeslcequivalent}, which we refer to as \emph{LC-optimization}.

Recent works have explored how to further optimize graph state preparation by employing two-qubit Clifford gates beyond simple $\CZ$ gates \cite{Gachechiladze2017,Patil2023,bataille2021reducedquantumcircuitsstabilizer,li2022photonic,kaldenbach2025efficientpreparationresourcestates,GoubaultdeBrugiere2025graphstatebased,davies2025preparinggraphstatesforbidding,kumabe2025complexitygraphstatepreparationclifford}.
As an example of such an optimization, consider the controlled-$X$ (controlled-NOT or $\mathrm{CNOT}$) gate, defined as $\CX_{i,j} = \ketbra{0}_i\otimes I_j + \ketbra{1}_i\otimes X_j$.
The action of this gate on a graph state can be expressed in terms of an equivalent set of $\CZ$ operations.
Recall the definition of $g_i$ \eqref{eq:graph_stabilizers}, and that $g_i\ket{G} = \ket{G}$.
This implies that 
\begin{equation}
    X_j\ket{G} = \prod_{k\in n(j)} Z_k\ket{G},
\end{equation}
and allows us to derive, for a pair of nodes $i,j$ not connected by an edge, the equivalence 
\begin{equation} \label{eq:cx_derivation}
 \begin{aligned}
    \CX_{i,j} \ket{G} &= \left(\ketbra{0}_i\otimes I_j\right) \ket{G} + \left(\ketbra{1}_i\otimes X_j\right) \ket{G} \\
    &= \left(\ketbra{0}_i\otimes I_j\right) \ket{G} + \left(\ketbra{1}_i\otimes \prod_{k\in n(j)}Z_k \right) \ket{G} \\
    &= \prod_{k\in n(j)} \left(\ketbra{0}_i\otimes I_j + \ketbra{1}_i\otimes Z_k\right) \ket{G} \\
    &=\prod_{k\in n(j)} \CZ_{i,k} \ket{G}
\end{aligned}   
\end{equation}
Therefore, a $\CX$ with control qubit $i$ and target $j$ has the same effect as a product of $\CZ$s between $i$ and each qubit in the neighborhood of $j$.
A $\CZ$ between two nodes complements the corresponding edge --- adding the edge if absent and removing it if present --- so a $\CX_{i,j}$ complements all edges between $i$ and the neighborhood of $j$, as shown in \cref{fig:main_rules}, middle.
If $n(j)$ contains more than one qubit, then this equation allows us to replace several gates with one, adding edges to a graph with fewer two-qubit gates than would be required if we used $\CZ$ gates alone.
A similar result can be derived for the gate $\CY$, defined here to be $|0\rangle\langle 0|\otimes I + |1\rangle\langle 1|\otimes ZX$ (note the missing factor of $i$ compared to the usual definition).
This gate is equivalent to $\CZ$ up to single-qubit unitaries, $\CY = (\sqrt{Z}\otimes \sqrt{X})\CZ(I\otimes \sqrt{X}^\dagger)$, and therefore has the same two-qubit gate cost.
However, we have the relation $\CY = \CZ \circ \CX$, leading to the identity
\begin{equation} \label{eq:rule_y}
    \CY_{i,j} \ket{G} = \prod_{k \in n(j) \cup \{j\}} \CZ_{i,k} \ket{G}\ .
\end{equation}
thereby adding or removing one more edge compared to \cref{eq:cx_derivation}, as shown in \cref{fig:main_rules}.
Given any two-qubit gate $U_{i,j}\in\{\CZ_{i,j},\CX_{i,j},\CY_{i,j}\}$, we can define the induced action $U_{i,j}(G)$ on the graph to be:
\begin{equation}
    \ket{U_{i,j}(G)} := U_{i,j}\ket{G}.
\end{equation}

\begin{figure}
    \centering
    \includegraphics[scale=0.23]{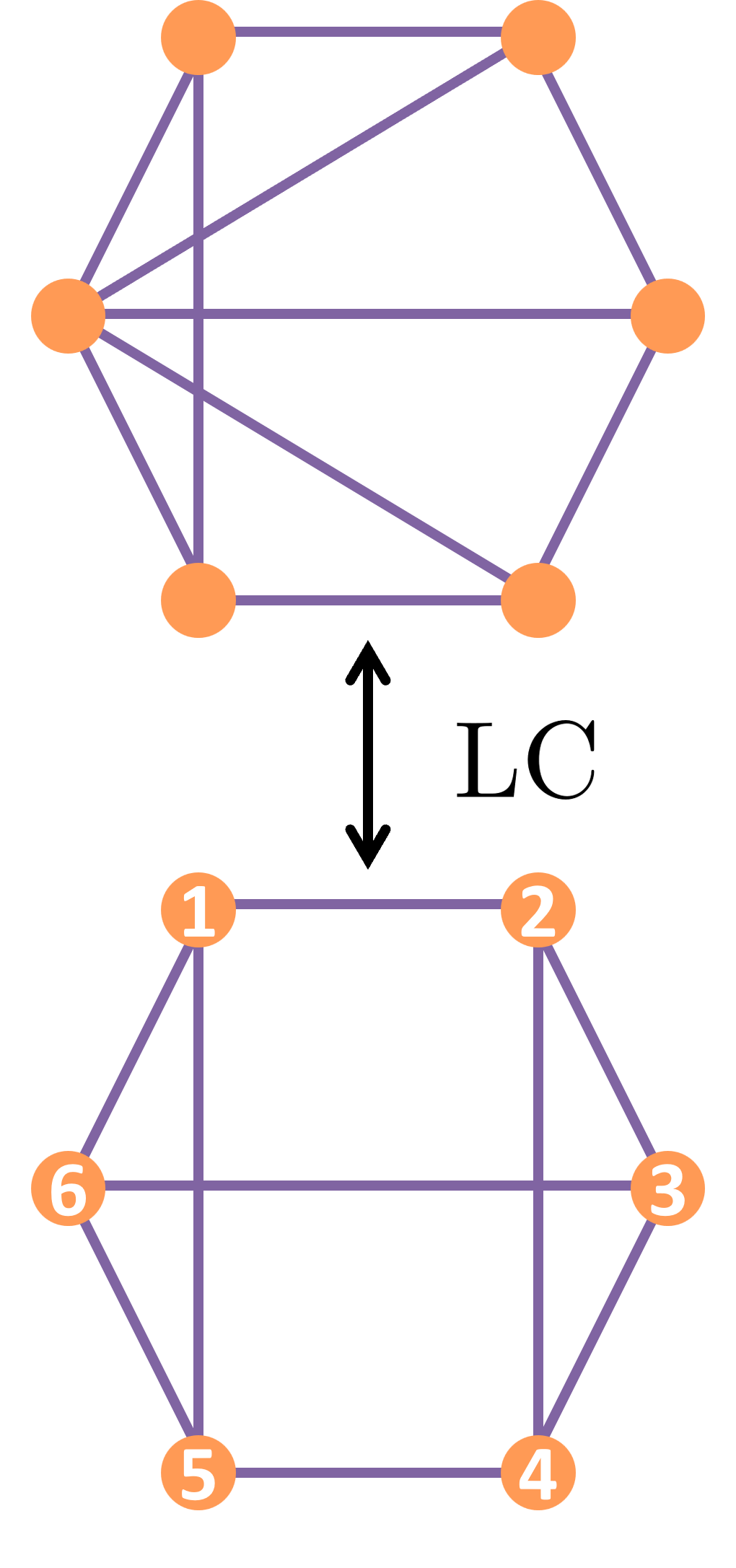}
    \hfill
    \raisebox{10em}{
    \scalebox{0.75}{
    \Qcircuit @C=.1em @R=1.5em @!R @!C
    {
    \lstick{|+\rangle_1} & \ctrl{1} & \qw & \qw & \qw & \qw & \ctrl{5} & \qw \\
    \lstick{|+\rangle_2} & \control \qw & \ctrl{2} & \qw & \targ & \ctrl{1} & \qw & \qw \\
    \lstick{|+\rangle_3} & \ctrl{1} & \qw & \ctrl{3} & \qw & \control \qw & \qw & \qw \\
    \lstick{|+\rangle_4} & \control \qw & \control \qw & \qw & \qw & \qw & \qw & \qw \\
    \lstick{|+\rangle_5} & \ctrl{1} & \qw & \qw & \ctrl{-3} & \qw & \qw & \qw \\
    \lstick{|+\rangle_6} & \control \qw & \qw & \control \qw & \qw & \qw & \control \qw & \qw
    }
    }
    }
    \hfill
    \raisebox{10em}{
    \scalebox{0.75}{
    \Qcircuit @C=.1em @R=1em @!R @!C 
    {
    \lstick{|+\rangle_1} & \ctrl{1} & \ctrl{4} & \qw & \qw & \qw  & \qw \\
    \lstick{|+\rangle_2} & \control \qw & \qw & \ctrl{2} & \qw & \qw &  \qw \\
    \lstick{|+\rangle_3} & \ctrl{1} & \qw & \qw & \qw & \ctrl{3} &  \qw \\
    \lstick{|+\rangle_4} & \control \qw & \qw & \gate{Y} & \ctrl{1} & \qw &  \qw \\
    \lstick{|+\rangle_5} & \ctrl{1} & \gate{Y} & \qw & \control \qw & \qw & \qw  \\
    \lstick{|+\rangle_6} & \control \qw & \qw & \qw & \qw & \control \qw & \qw
    }
    }
    }
    \caption{Left: Example of a pair of graphs that define a complete orbit under local complementation (LC), up to node relabeling.
    The minimal edge count in the LC orbit is thus equal to 9 (bottom graph).
    Middle: Circuit derived using our technique which uses a $\CX$ gate to prepare the bottom-left graph state using 8 two-qubit gates.
    Right: Circuit that uses $\CY$ gates to prepare the same graph state using 7 two-qubit gates.}
    \label{fig:6_qubit_example}
\end{figure}

Using \cref{eq:cx_derivation,eq:rule_y}, one can reduce the number of two-qubit gates needed to construct a given graph state by finding an appropriate sequence of $\CX$, $\CY$ and $\CZ$ gates that generate all necessary edges. This method can produce circuits with lower two-qubit gate count compared to what is possible with LC-optimization. As an example, \cref{fig:6_qubit_example} (left) shows a graph with the property that its LC-orbit contains only two distinct graphs, up to relabeling of nodes. The smaller edge count of the two graphs is $|E| = 9$, meaning that the minimal two-qubit gate count achievable with LC-optimization is 9. In contrast, \cref{fig:6_qubit_example} (right) shows a circuit constructed using \cref{eq:cx_derivation,eq:rule_y} with 7 two-qubit gates.
In practice, it may be helpful to use the two methods in tandem, first reducing edge count with LC-optimization before synthesizing the optimized graph with $\CX$, $\CY$, and $\CZ$ gates. 
It is also possible to alternate between $\CZ$ gates and $\LC$ operations, as explored in Ref.~\cite{davies2025preparinggraphstatesforbidding,jena2024graph}.
The circuits studied here are of this type, since we have $\CY_{i,j}\ket{G} = \LC_j \circ \CZ_{i,j}\circ  \LC_j\ket{G}$ and $\CX_{i,j}\ket{G} = \LC_j\LC_k\LC_j \circ\CZ_{i,j}\circ \LC_j\LC_k\LC_j\ket{G}$ for any $k\in n(j)$.

\begin{figure}
    \centering
    \includegraphics[scale=0.23]{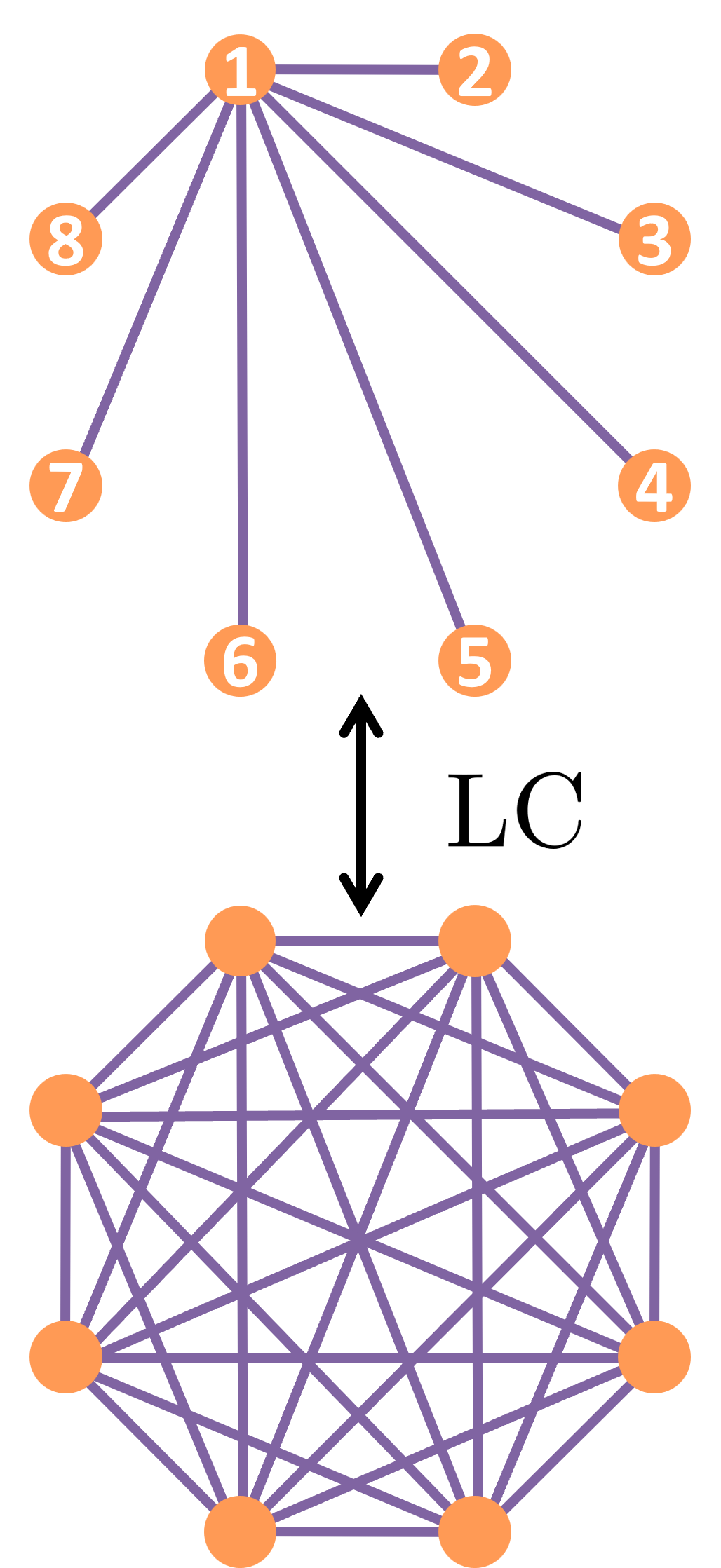}
    \hfill
    \raisebox{11em}{
    \scalebox{0.8}{
    \Qcircuit @C=.1em @R=1em @!R @!C
    {
    \lstick{|+\rangle_1} & \ctrl{1} & \qw & \qw & \qw & \qw & \qw & \qw & \qw\\
    \lstick{|+\rangle_2} & \control \qw & \targ & \qw & \qw & \qw & \qw & \qw & \qw\\
    \lstick{|+\rangle_3} & \qw & \ctrl{-1} & \targ & \qw &  \qw & \qw & \qw& \qw \\
    \lstick{|+\rangle_4} & \qw &  \qw & \ctrl{-1} & \targ & \qw & \qw & \qw & \qw\\
    \lstick{|+\rangle_5} & \qw & \qw & \qw & \ctrl{-1} & \targ & \qw & \qw & \qw\\
    \lstick{|+\rangle_6} & \qw & \qw &  \qw & \qw & \ctrl{-1} &  \targ & \qw & \qw\\
    \lstick{|+\rangle_7} & \qw & \qw &  \qw & \qw & \qw &  \ctrl{-1} & \targ & \qw\\
    \lstick{|+\rangle_8} & \qw & \qw &  \qw & \qw & \qw &  \qw & \ctrl{-1} & \qw
    }
    }
    }
    \hfill
    \raisebox{11em}{
    \scalebox{0.8}{
    \Qcircuit @C=.1em @R=1em @!R @!C 
    {
    \lstick{|+\rangle_1} & \ctrl{4} & \ctrl{2} & \ctrl{1} & \qw \\
    \lstick{|+\rangle_2} & \qw & \qw & \control \qw & \qw \\
    \lstick{|+\rangle_3} & \qw & \control \qw & \targ & \qw  \\
    \lstick{|+\rangle_4} & \qw &  \qw & \ctrl{-1} & \qw \\
    \lstick{|+\rangle_5} & \control \qw & \targ & \targ & \qw \\
    \lstick{|+\rangle_6} & \qw & \qw &  \ctrl{-1} & \qw  \\
    \lstick{|+\rangle_7} & \qw & \ctrl{-2} &  \targ & \qw \\
    \lstick{|+\rangle_8} & \qw & \qw &  \ctrl{-1} & \qw 
    }
    }
    }
    \caption{Left: The star and fully connected graphs define an LC-orbit representing the GHZ state.
    Middle: A circuit to prepare the star graph using $\CX$ gates that simplifies connectivity to nearest-neighbor.
    Right: A circuit that reduces depth from linear to logarithmic in system size.}
    \label{fig:star}
\end{figure}

Beyond reducing gate count, using $\CX$ and $\CY$ gates can also reduce circuit depth and simplify circuit connectivity.
A simple example is given by the $N$-qubit star graph, which is equivalent up to local Cliffords to the GHZ state, as shown in \cref{fig:star}.
The only other graph in the LC-orbit of the star graph is the fully connected graph.
The maximum degree in both cases is $\delta = N$, such that creating the state using $\CZ$ gates requires both long-range gates and a depth proportional to system size.
By using $\CX$ gates, we can either prepare the state using nearest-neighbor gates only, or with circuit depth logarithmic in system size, see \cref{fig:star}.
In this case, this reproduces well-known circuits for preparing the GHZ state, but the same insights can be applied more generally.

\subsection{Graph decimation}
\label{sec:graph_decimation}

The graph state approach provides a powerful framework for the application of modern machine learning techniques to state preparation. While for small codes gates can be intuitively applied using simple transformation rules (e.g., the circuits in \cref{fig:6_qubit_example,fig:star} were derived by hand), for larger codes it is critical to automate the procedure.
In this work, we consider the task of finding a minimal sequence of gates that disentangles a given graph state (disconnecting all nodes in the graph), which we refer to as \emph{graph decimation}. Once the graph is disentangled, we can simply reverse the order of the gate sequence to produce a circuit that prepares the target graph state from the all-$\ket{+}$ product state. 

\begin{figure}[h!]
\includegraphics[width=8cm]{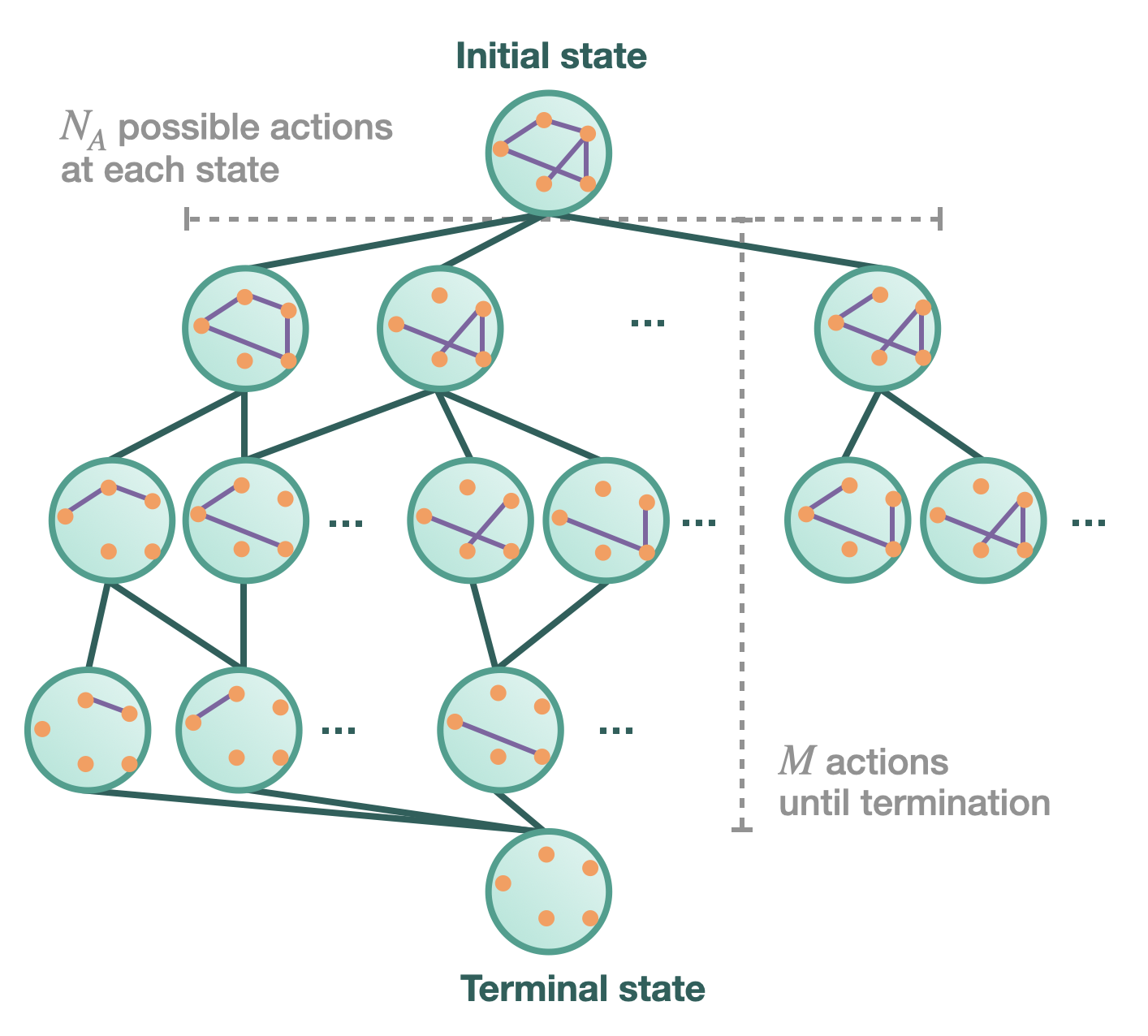}
\caption{Schematic of the graph decimation task. The initial state is the graph of the quantum state we seek to implement. The terminal state is the fully disconnected graph with no edges.} 
\label{fig:tree_of_states}
\end{figure}

We illustrate graph decimation as a sequential decision-making problem in \cref{fig:tree_of_states}. At each step, we choose a particular two-qubit gate to apply to the graph state. Applying a gate corresponds to taking an action on a graph that can disconnect or reconnect nodes, and the process ends at the fully disconnected graph. The choice to represent the state in terms of a graph provides a clear metric---the number of edges---that indicates progress towards a solution, and provides a simple way to rank all possible gates that may be applied at each step. Furthermore, in cases where many gates remove the same number of edges, additional metrics such as depth or connectivity of the graph can be considered as `tie-breakers', as in the example circuits in \cref{fig:star} which have the same number of gates but different connectivity and circuit depth. This gives the approach a high degree of flexibility and allows it to simultaneously optimize different circuit metrics.

The fact that graph decimation works ``in reverse'' has algorithmic advantages. Starting from the original graph presents our algorithms with the most powerful gates at the start of the procedure ($\CX$ or $\CY$ gates targeting high-degree nodes). For example, consider the following circuit that performs graph decimation on a 4-cycle,
\begin{equation} \label{eq:4cycle}
\adjustbox{valign=m}{\includegraphics[scale = 0.23]{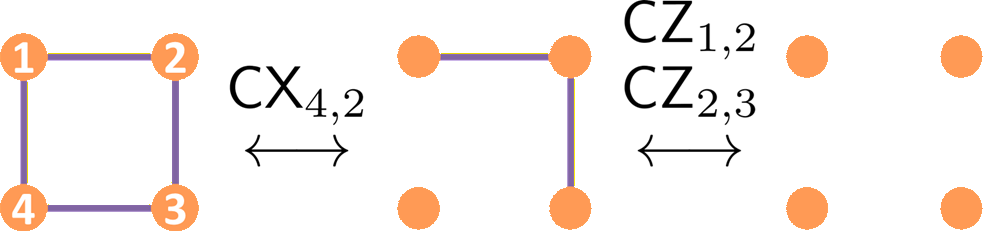}}
\end{equation}
When going from left-to-right, as in graph decimation, it is easy to identify an advantageous $\CX$ that removes two edges, and then apply the remaining $\CZ$s to ``clean up'' the remaining edges. Conversely, if we were to go from right-to-left, we would need to first ``set up'' the $\CX$ by performing two appropriate $\CZ$s sharing a node, which requires some foresight. If we were to instead start by applying two disjoint $\CZ$s, there would be no opportunity for savings.

Graph decimation is a highly non-linear process, in that the effects of a given gate at any step in the process depend on the current graph structure, which itself depends on the previously applied gates, as seen by the 4-cycle example. Therefore, gates which are locally optimal, removing the highest number of edges among all gates at a given step, may in fact be globally suboptimal. Furthermore,
as we derive explicitly in \cref{sec:combinatorial_space}, the space of possible actions in an exhaustive search solution of the graph decimation problem scales as $O(N_Q^{2M})$ for codes with $N_Q$ qubits and a circuit of $M$ total gates. Already for the $23$-qubit Golay code, this corresponds to a solution space size $\sim 10^{136}$. Clearly, efficient search and optimization techniques are critical to find good solutions to the graph decimation problem. These are the focus of the rest of this work.

\section{Heuristic search methods}
\label{sec:heuristic_method}

As explained in the previous section, a full search among all possible combinations of actions to find the optimal solution is infeasible even for small codes.
Therefore, solving the problem necessitates more sophisticated methods.
Before introducing the reinforcement learning methods that represent the main results of this work, we extensively benchmark the performance of heuristic search approaches for graph decimation. We describe best-first search (BeFS) in \cref{sec:BeFS} and its generalization to beam search in \cref{sec:beam search}.
The best solutions provided by all methods are compared in \cref{sec:results}, while the corresponding time to solution comparison is shown in \cref{sec:time_comparisons}.

\begin{figure*}[htp]
\centering
\includegraphics[width=\linewidth]{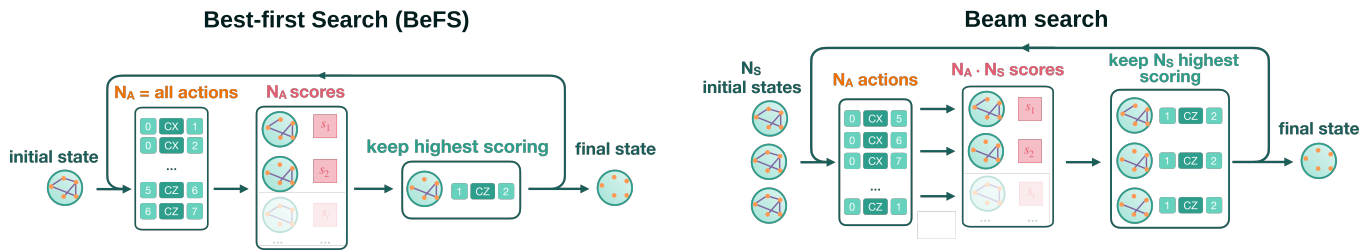}
\caption{Schematics of the heuristic search methods developed in this work. Left: best-first search (BeFS); Right: Beam search algorithm.
}
\label{fig:methods} 
\end{figure*}

\subsection{Best-first search (BeFS)}
\label{sec:BeFS} 

A greedy best-first search (BeFS) algorithm serves as a benchmark solution method. In BeFS, all possible actions on the current state are considered, and the next state is selected based on a heuristic scoring function, with the highest-scoring action added as the next two-qubit gate in the circuit. The process of expansion and selection is then repeated for the next state. \Cref{fig:methods} (left) illustrates the BeFS algorithm loop.

Since the objective of the task is to disconnect the graph in as few actions as possible, we use as a scoring function the number of edges removed by an action, plus a small (tunable) negative score if the action added a new circuit layer. However, many actions remove the same number of edges from a given state in a step of the decimation procedure, but can result in final circuits with different two-qubit gate counts, due to different subsequent action possibilities. In order to further rank these actions to allow an overall lower gate count, we adopted tie-breaking heuristics, including \textit{min-degree} tie-breaking (preferring actions that remove edges from low-degree nodes) and \textit{random} tie-breaking. 
By systematically applying these heuristics, we found that the \textit{random} tie-breaker was best for small codes, as it allows the largest space of solutions to be explored, while the \textit{min-degree} tie-breaker resulted in better solutions on large codes, because the graph tends to be disentangled sequentially while preserving high-degree nodes which, as discussed in Section \ref{sec:theory}, provide the greatest opportunity for gate savings. Even when using the \textit{min-degree} tie breaker, some actions are still ranked equivalently, in which case we resort to random tie breaking. This randomness makes the algorithm non-deterministic, such that we can repeat it many times to improve the solution quality. This repetition constitutes a majority of the runtime of BeFS. 

BeFS has space complexity $O(N_A^{\mathrm{tot}})$ and time complexity $O(N_A^{\mathrm{tot}} M)$, where $N_A^{\mathrm{tot}}$ is the total number of gates that can be applied at any given step, as computed in \cref{eq:NA}, and $M$ is the sequence length of the solution. A natural extension of BeFS is to ``look ahead'', retaining multiple states at each step and trading increased complexity for improved accuracy. This is the core idea of the beam search algorithm, discussed in the next section.

\subsection{Beam search}
\label{sec:beam search}
In contrast to the single state selected at each step in BeFS, beam search selects the top $N_S$ states at each step, with $N_S$ referred to as the \emph{beam width}. \Cref{fig:methods} (right) illustrates the beam search algorithm. Each state is expanded by $N_A$ randomly sampled actions (with $1 \leq N_A \leq N_A^{\mathrm{tot}}$, $N_A^{\mathrm{tot}}$ being the maximum number of possible actions as computed in \cref{eq:NA}), resulting in $N_S \cdot N_A$ intermediate states. These are ranked by score and the best $N_S$ 
are retained for the next iteration.

It is important to note that, while the number of possible actions at each step scales quadratically with the number of qubits, calculating $N_A$ new states for each of the $N_S$ states retained from the previous iteration may still be impractical due to memory constraints. 
For the states considered in this work (see \cref{table_results}), maintaining high beam width at the cost of random action sampling results in the most efficient circuits.
Moreover, repeating this randomized beam search for many iterations effectively explores a larger set of actions, while a non-randomized beam search would always explore the same small set of states.

Beam search is equivalent to BeFS with random tie-breaking when $N_S = 1$ and $N_A = N_A^{\mathrm{tot}}$, i.e. all actions are considered. The flexibility to configure $N_S$ and $N_A$ gives beam search two advantages over BeFS. Firstly, $N_S$ > 1 allows states to be explored that may have locally suboptimal scores at some steps but higher cumulative scores (and consequently better circuits) overall. Secondly, when $N_A < N_A^{\mathrm{tot}}$, i.e. strictly smaller than the maximum, there is a chance that some locally optimal but globally suboptimal actions will be randomly excluded from the set of considered actions. This allows further exploration of the state space by excluding high-scoring actions that may ``crowd-out'' other lower-scoring but ultimately better actions, as discussed in \cref{{sec:beam_search_params}}. 

Beam search also has disadvantages; the large number of intermediate states causes memory requirements to scale as $O(N_S N_A)$ and time complexity as $O(N_S N_A M)$. However, the algorithm is trivial to parallelize, and we were able to implement it using JAX \cite{jax2018github} to leverage accelerator hardware with large memory capacity. We discuss the details and impact of our JAX implementation further in \cref{sec:time_comparisons}.

Our beam search implementation has three additional features that improve the efficiency of the search. Firstly, when one branch of the search terminates, we restart the algorithm,
since the other branches cannot improve on the number of two-qubit gates of the current best solution. Secondly, the number of edges removed at each step is normalized by the number of edges of the corresponding state at the current step. This ensures the scores of states grow in magnitude as the graph gets more disconnected, therefore graph states closer to termination are more likely to be selected at each step. 
Finally actions can be ``masked'' to prioritize those that remove one or more edges. Action masking is discussed further in 
\cref{sec:beam_search_params}. 

\section{Search-informed reinforcement learning with \textit{QuSynth}}
\label{sec:ml_methods}

To go beyond methods based on heuristic search and find more efficient solutions for the state preparation problem, in this section we describe \textit{QuSynth}, a hybrid method that uses Monte Carlo tree search (MCTS) to augment the planning of a reinforcement learning agent. We first discuss Monte Carlo tree search before moving to a discussion of reinforcement learning. We also describe a standard model-free reinforcement learning agent, which we later compare against \textit{QuSynth}.

\subsection{Monte Carlo tree search}
\label{sec:mcts}

\begin{figure}[htp]
\centering
\includegraphics[width=\linewidth]{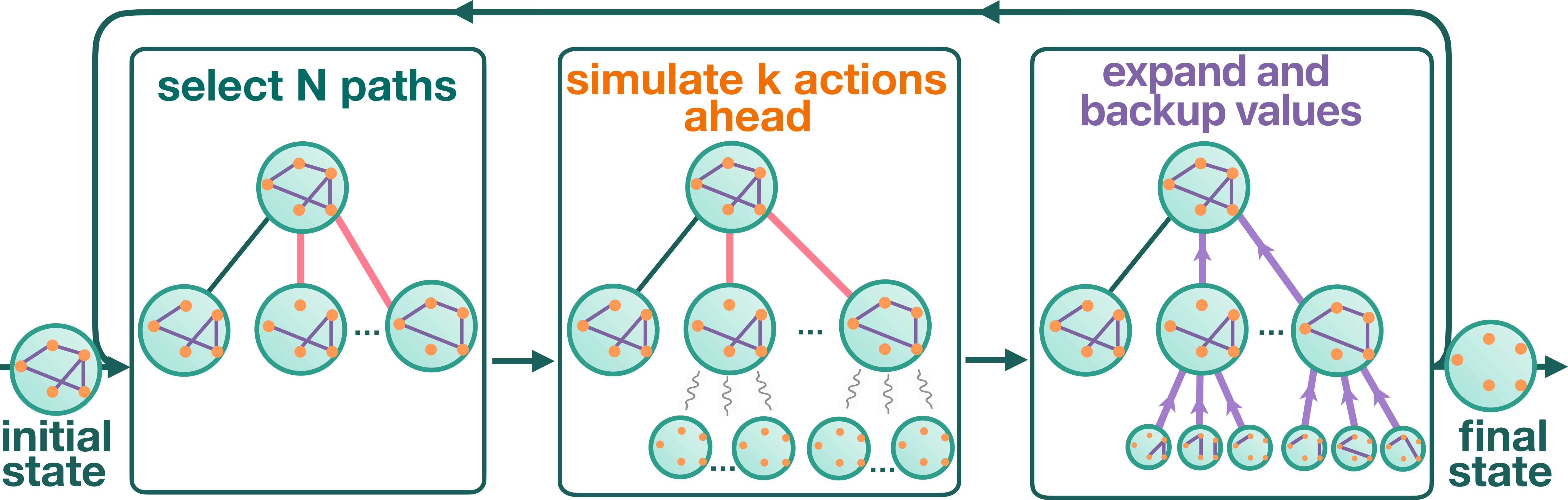}
\caption{Schematic of Monte Carlo tree search. Promising states for further exploration are identified, and simulations project the reward from $k$ actions from each state. New states are added to the tree and projected values from the simulation phase are backpropagated through the tree.
}
\label{fig:mcts} 
\end{figure}

Monte Carlo tree search (MCTS) is a planning algorithm that is widely used for decision making in large combinatorial search spaces.
It excels especially for problems where local heuristics are not able to reliably predict the value of an action, and actions that are apparently suboptimal may be part of the best global solutions \cite{browne2012mctsreview}. While MCTS is, at its core, a sophisticated classical search algorithm, it is also amenable to agent guidance in reinforcement learning, as we discuss in the next section.

The basis of MCTS is to incrementally build a search tree by iteratively performing the following four steps (see also \cref{fig:mcts}):
\begin{enumerate}
    \item \textit{Selection}: the algorithm traverses the current tree according to a policy that balances `exploring' under-visited states and `exploiting' high-quality states, and identifies the most promising state for further exploration.
    \item \textit{Expansion}: new actions are taken from that state and the resulting states are added as new nodes in the tree.
    \item \textit{Simulation}: the value of new nodes is estimated. This estimate is traditionally based on the outcome of taking random actions from that state to the terminal state (in our case the terminal state is a graph with no edges remaining). In agent-guided search, the state value is predicted by the agent.
    \item \textit{Backup}: simulation results are used to assign scores to each new node, and to update the expected rewards and visit counts of all intermediate states that lead to the new node(s).
\end{enumerate}
Over repeated iterations, MCTS concentrates search on the most promising regions of the space, allowing efficient identification of optimal or near-optimal solutions.

The core functionality of the algorithm hinges on how promising nodes are selected for further exploration. Taking action $a$ from state $s$ yields a new state $s'$. A state is typically selected for further exploration based on its score using the ``Upper Confidence Bound for Trees'' (UCT) formula \cite{kocsis2006uct}
\begin{equation}
    \text{UCT}(s') = \bar{Q}(s') + c \cdot P(s, a) \cdot \frac{N(s)}{1 + N(s')}.
\label{eq:UCT}
\end{equation}
$\bar{Q}(s')$ is the average cumulative reward or `quality' of games involving the state $s'$ that have occurred so far in the tree search. The quality of any simulation involving $s'$ is stored in the tree during the backup phase. 
$P(s, a)$ is the prior probability of taking action $a$ from $s$. In agent-guided MCTS, this is predicted by the agent; otherwise, it may be uniform or set using a heuristic.
$N(s)$ and $N(s')$ are the number of times state $s$ and $s'$ have each been visited so far during the tree search. 
Roughly speaking, the first term favors high-scoring nodes, while the second term favors nodes that have not been explored extensively. The two are balanced by the (non-negative) hyperparameter $c$ that controls the preference for exploitation of high-quality states (preferred when $c$ is near zero) over exploration of states with few previous visits (preferred when $c$ is large).

In its standard implementation, MCTS is a sequential algorithm where the selection, expansion, and simulation are done for a single promising path and the tree statistics are updated before selecting a new promising path.
However, to enable batch computations and efficient GPU usage, for the results in this work we find it beneficial to instead consider simultaneously the $k$ best states in the tree and evaluate $m$ new actions from each. This allows us to update batches of $k \cdot m$ states in parallel. For batches of trees, we perform the selection, expansion, and backup phases sequentially but perform the simulation phase on the full batch simultaneously.

\subsection{Reinforcement Learning framework}
Unlike heuristic search methods, reinforcement learning addresses sequential decision-making problems by specifying a reward objective and learning how actions influence long-term returns. It aims to maximize future rewards by learning a policy that, at each step, can be used to select an action in response to an observation of the state of an environment. In a typical setting, a neural network learns the policy and also predicts the value of intermediate states even before the final result of taking a sequence of actions is known. A separate \emph{policy network} and \emph{value network} might be used, which is the case for the model-free reinforcement learning model we consider, or a single network with a policy head and value head can be employed instead, which is what we do for the MCTS-guided reinforcement learning implementation we explore. The policy is trained by simulating a sequence of actions during ``self-play''. Each state encountered is assigned a score either based on the final outcome or based on a prediction by the value network. During training, these scores are used update the policy. 

In \cref{sec:results}, we compare the performance of a standard model-free reinforcement learning agent with that of a hybrid method using tree search to guide the reinforcement learning agent. For both methods, we use the upper triangle of the adjacency matrix as a state observation, in addition to the qubits on which a gate acts in the current circuit layer (for model-free RL only), and the number of edges removed by actions from that state (for MCTS-guided RL only). Based on the state observation, the agent chooses the next action and the state is advanced by that action. In model-free reinforcement learning, the next action is predicted directly from the agent policy, while in MCTS-guided reinforcement learning the agent policy and state value predictions inform the tree search that eventually decides the next action. 

For both reinforcement learning methods the agent is a multi-layer perceptron (MLP) with two hidden layers, each ranging from 64 units for the 23-qubit Golay code to 512 units for the largest codes. In the model-free context we also considered graph neural networks for the agent architecture but found them not to improve over the MLP, perhaps due to the increasing sparsity of the graph structure \cite{gnn_sparsity_degradation_2023}. 
To make the size of the action space more tractable, we prevent the agent from choosing actions that do not remove at least one edge. This reduces the action space by almost an order of magnitude for some codes (see \cref{sec:combinatorial_space}) and ensures that episodes always terminate.

Since reinforcement learning agents generate their own training data during self-play, the computational efficiency of simulation is an enormous advantage for effective training. In this sense, the model-free reinforcement learning has an advantage because we implement the environment in JAX \cite{jax2018github} and compile the training loop as a single program that can be run on GPU and parallelized to thousands of environments. The tree search in MCTS-guided reinforcement learning allows more planning since it has a built-in method for the agent to look ahead to states several actions in the future when choosing an action. However, the tradeoff is that MCTS is less amenable to massive GPU parallelization and the agent-guided tree search requires substantially more compute resources per move. Indeed, we find that MCTS-guided reinforcement learning generally finds more optimal solutions, but at the expense of much larger computational resources. Detailed results for all codes are shown in \cref{sec:results}.

\subsubsection{Model-free reinforcement learning}

\begin{figure}[htp]
\centering
\includegraphics[width=\linewidth]{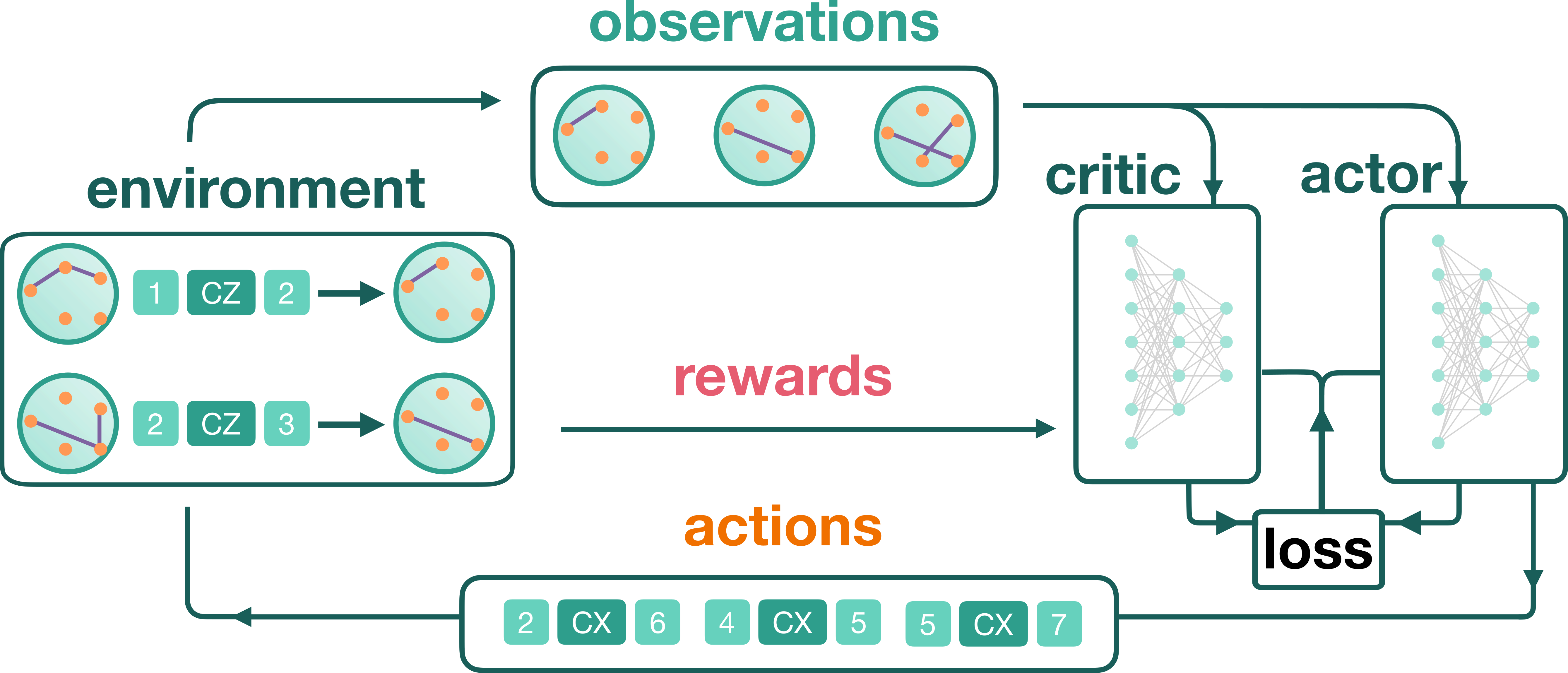}
\caption{Schematic of the training loop for model-free reinforcement learning. The agent makes observations of the graph states and chooses actions to take, then updates the states by taking those actions. Rewards are used for training the agent.
}
\label{fig:RL} 
\end{figure}

\begin{figure*}[htp]
\centering
\includegraphics[width=0.85\linewidth]{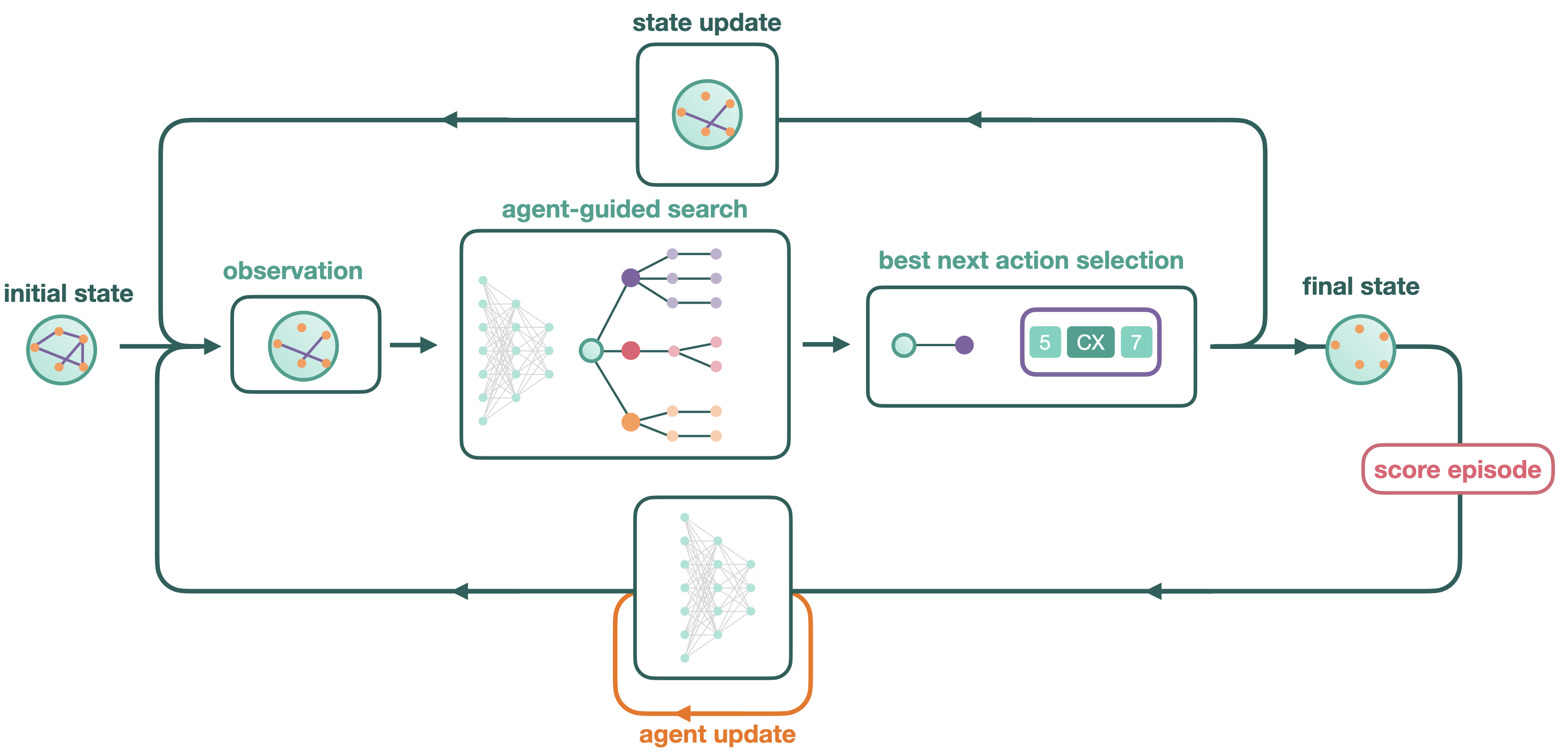}
\caption{Schematic of \textit{QuSynth} (MCTS+RL) method. The initial state is observed by the agent, which is used to guide a tree search to act on the graph state for a few steps. The most visited action is then selected as best action and used to update the state.  The loop continues until the final state is reached. At the end of a trajectory, the corresponding score is computed to evaluate the quality of the result, and it is used to update the neural-network agent.}
\label{fig:mQuSynth_method} 
\end{figure*}

We train the model-free RL agent with Proximal Policy Optimization (PPO) \cite{schulman2017proximalpolicyoptimizationalgorithms}, which has been used in successful applications of RL in challenging domains \cite{openai2019dota2largescale}. The PPO loss function for a target policy $\pi$ is
\begin{equation}
L_{\text{PPO}} = c_{\pi} L_{\pi} + c_{V} L_{V} + c_{H} H(\pi)
\label{eq:ppo_loss}
\end{equation}
The value loss $L_V$ is the mean squared error between the predicted state values and the empirical returns. The policy loss $L_\pi$ encourages actions that yield higher-than-expected returns by maximizing the probability of advantageous actions, while stabilizing the training by preventing large single policy updates. The entropy term $H(\pi)$ measures the randomness of the policy distribution $\pi$; we take its coefficient $c_H$ to be negative to maximize entropy, which encourages exploration and prevents premature convergence to suboptimal deterministic policies. We discuss the empirical results of the interaction of these loss components in \cref{sec:rl_training} and show the optimized hyperparameters in \cref{table:hyperparam_results}.

A clear disadvantage of the RL approach described here is that it necessitates training separate agents for each code. In \cref{sec:transformer_bc} we describe a method to train a generalist graph state preparation agent using behavior cloning with a Transformer. Despite successful behavior cloning we found that this agent was still not able to generalize its performance to unseen codes, but could provide a basis for future work in this direction.

\subsection{\textit{QuSynth}: MCTS-guided reinforcement learning}

Reinforcement learning can also be used in tandem with Monte Carlo tree search to enable more sophisticated planning and decision making.
This type of approach became famous as the core of \textit{AlphaZero} \cite{alphazero}, where an agent-guided tree search beats expert human players at complex games. As opposed to the agent choosing actions directly from its policy as in the previous section, here the policy and state value predictions of the agent guide a tree search, and the most popular first action in the tree search is chosen as the action in self-play. 

In comparison to MCTS without reinforcement learning, we emphasize several key differences. Here the value of a new state $Q(s')$ and the action prior $P(s')$ in \cref{eq:UCT} are predicted by the agent. This completely removes the random simulation phase of traditional MCTS. We also emphasize that here, MCTS only serves as an engine for choosing the best next move from the current state during self-play. In traditional MCTS a tree may represent a full game: its root represents the initial graph and its eventual depth can be as large as the number of actions to finish the game. Here we use comparatively small trees (depth of $5$ to $10$) that are rooted at the current state in self-play. The tree search therefore provides policy guidance based on a 5--10 move look-ahead in the game.

Once the terminal state of the game is reached at the end of an episode (the graph is fully disconnected), each intermediate state of the game that was encountered during the episode receives a score that is the number of actions the agent took between that state and the end of the game.
The agent is trained to predict this score $y$ and the MCTS policy $\pi$ of each state. The loss function for the agent is
\begin{equation}
    L = c_V \text{MSE}(y,\hat{y}) + c_\pi \text{CE} (\pi,\hat{\pi}) + c_\text{KL} \text{KL}(\hat{\pi}, \pi_\text{ref}),
\end{equation}
where $(\hat{y},\hat{\pi})$ are the score and policy predicted by the agent and $\pi_\text{ref}$ is a reference policy from the heuristic. The mean squared error (MSE) loss encourages the agent to correctly predict the state score, the cross entropy (CE) loss encourages the agent to correctly predict the target (MCTS root) policy. The KL divergence term between the proposed policy and a reference policy helps to stabilize the training by preventing dramatic policy changes. 

For stable training and self-play, we initially tether the state value and policy predictions from the network to heuristic values, and gradually reduce the contribution of the heuristic. We use the sum of edges removed by all actions from a state as a heuristic for the state value. We generally find that the value prediction trains more quickly and has more positive influence on the agent performance than the policy prediction.

\section{Results}
\label{sec:results}

\begin{table*}[t]
\small
\setlength{\tabcolsep}{8pt}
\renewcommand{\arraystretch}{1.2}
\definecolor{tealrow}{HTML}{E1F6F2}
\resizebox{\textwidth}{!}{
\begin{tabular}{c | c  c | c c c c }
\toprule
\textbf{Code} & Graph edges & Baseline & \textbf{BeFS} & \textbf{Beam search} & \textbf{RL} & \textbf{\textit{QuSynth}} \\
  &  & (TQGs, TQd) & (TQGs, TQd) & (TQGs, TQd) & (TQGs, TQd) & (TQGs, TQd) \\
\midrule
\rowcolor{tealrow} $\nkd{9}{1}{3}$ Surface code  &  10 & (8, 3)$^1$  & (8, 3)$^a$ & - & - & - \\ 
$\nkd{12}{2}{4}$ Carbon code  & 21 & (16, 5)$^1$  & (16, \textbf{4})$^a$ & - & - & - \\ 
\rowcolor{tealrow} $\nkd{15}{1}{3}$ Reed-Muller code & 32 &  (23, 5)$^1$  & (23, \textbf{4})$^a$ & - & - & - \\
$\nkd{15}{7}{3}$ Hamming code  & 28 & (22, 4)$^1$  & (22, 4)$^a$ & - & - & - \\ 
\rowcolor{tealrow} $\nkd{17}{1}{5}$ Color code & 32 &  (23, 4)$^1$  & (23, 4)$^a$ & - & - & - \\ 
$\nkd{19}{1}{5}$ Color code & 41 & (27, 4)$^1$  & (27, 4)$^a$ & - & - & - \\ 
\rowcolor{tealrow} $\nkd{23}{1}{7}$ Golay code \cite{steane1996} & 77 & (56, 17)$^2$  & (46, 9)$^a$ & (45, \textbf{7}) & (45, \textbf{7}) & (\textbf{44}, 8) \\ 
$\nkd{36}{8}{6}$   &  158 & (124, 11$^{+1}$)$^3$  & (84, 13)$^a$ & (78, \textbf{10}) & (81*, 12) & (\textbf{77}, 11) \\ 
\rowcolor{tealrow} $\nkd{40}{20}{6}$ (non-CSS) & 396 & (284, 21$^{+1}$)$^3$ & (142, 22)$^b$ & (136, 31) & (132, \textbf{21}) & (\textbf{130}, 24) \\
$\nkd{42}{12}{5}$  & 193 & (131, 13$^{+1}$)$^3$  &  (95, 16)$^b$ & (\textbf{90}, 22)$^b$ & (\textbf{90}, 11) & (\textbf{90}, \textbf{9}) \\ 
\rowcolor{tealrow} $\nkd{48}{6}{8}$ & 231 & (203, 15$^{+1}$)$^3$  & (121, 30)$^b$   &  (117*, 18) & (117, \textbf{13}) & (\textbf{116}, 15) \\ 
$\nkd{64}{12}{8}$  & 462 & (298, 19$^{+1}$)$^3$  & (193, 32)$^b$  & (177*, 14) & (174, 16) & (170, 18) \\ 
 + LC-optimization & 298 & & - & - & (165, 21)$^\ddagger$ & (\textbf{161}, \textbf{13})$^\ddagger$ \\
\rowcolor{tealrow} $\nkd{80}{8}{10}$ & 524 & (446, 21$^{+1}$)$^3$  & (250, 30)$^b$  & (252, 22) & (251, 28) & (252, \textbf{13}) \\ 
\rowcolor{tealrow} + LC-optimization & 446 & & - & - & (\textbf{240}, 31)$^\ddagger$ & (245, 21)$^\ddagger$ \\
$\nkd{144}{12}{12}$ Gross code \cite{Bravyi2024} & 2192 & (876, 37$^{+1}$)$^3$ & (482, 46)$^b$ &  (427, 30)  & (413, 28) & (426, 26) \\ 
+ LC-optimization & 876 &   & (429, 55)$^{b, \ddagger}$  &  (381, 50)$^\ddagger$  & (\textbf{357}, \textbf{18})$^\ddagger$ & (362, 21)$^\ddagger$ \\
\bottomrule
\end{tabular}
}
\caption{Two-qubit gate counts (TQGs) and two-qubit circuit depths (TQd), respectively, of circuits obtained with different optimization techniques for the $\ket{0}_L$ ($\ket{+}_L$ for the Reed-Muller code) state of various stabilizer codes.
The first column lists stabilizer codes with $\nkd{n}{k}{d}$ notation and corresponding name (if present). The second one provides the initial number of edges for the corresponding graph. Previous best literature results (when available) and values obtained using the state preparation described in \cref{circuits} are shown as a reference in the third column. In the following columns we show the results obtained with the methods developed in this work: best-first search (BeFS), beam search, model-free reinforcement learning (RL) and \textit{QuSynth}. We did not use the most sophisticated methods for codes smaller than the Golay code, since it is already enough to beat or match the state of the art. The depth for the codes larger than the Golay code is obtained by the maximum degree of the graph: the $^{+1}$ indicates that the depth obtained could be higher (by 1) than the minimum value shown; the true depth depends on solving a graph coloring problem. The asterisk (*) denotes results improved by post-hoc compilation. The upper indices $a$ and $b$ signal results obtained with \textit{random} and \textit{min-degree} tie-breakers, respectively. ($^1$)  refers to the results in \cite{peham2024automated}, ($^2$) refers to the result in \cite{webster2025}, ($^3$) refers to those obtained using only $\CZ$ gates applying the state preparation described in \cref{circuits} with LC optimization. ($^\ddagger$) refers to results obtained applying the decimation to an LC-optimized graph.
} 
\label{table_results}
\end{table*}

In this section, we demonstrate the performance and scalability of \textit{QuSynth} and the other methods discussed in \cref{sec:heuristic_method,sec:ml_methods} when applied to the preparation of logical stabilizer states of QEC codes. These codes range from $n=9$ to $n=144$ qubits, include both Calderbank–Shor–Steane (CSS)~\cite{steane_1996, calderbank_1996} codes and non-CSS codes, and have been selected to provide a representative range of physical qubit requirements and code distances. The parameters $\nkd{n}{k}{d}$ of the selected codes are shown in the first column of \cref{table_results}. To obtain the initial graph, which is the input to the graph decimation procedure, we use the \textit{Stim}~\cite{gidney2021stim} implementation of the Gaussian elimination procedure that converts an arbitrary stabilizer state into a graph state with local Clifford decorators (see \cref{eq:stabilizer_to_graph} in \cref{sec:theory}). Notably, for CSS states, the initial graph is bipartite, which means that the circuits resulting from graph decimation can be re-written in terms of \CX gates acting on initial $\ket{0}$ and $\ket{+}$ states (see \cref{app:bipartite_graphs}). 

The results for each code were obtained using an NVIDIA A100 GPU with 80GB of memory. We made efforts to optimize the hyperparameters of each method to improve results. We ran BeFS for $10^6$ iterations. Beam search was run for each code until no further improvement in results was observed, and we tuned the beam width and the number of actions $N_A$ (see \cref{sec:beam_search_params}). Model-free RL is trained until policy convergence (see \cref{sec:rl_training}), and \textit{QuSynth} (MCTS+RL) is trained until the episode rewards saturate. To automatically optimize the many hyperparameters (e.g. learning rate, PPO discount factors, entropy and value coefficients) for the model-free RL, we used the \textit{PROTEIN} algorithm from the \textit{PufferLib 3.0} library \cite{PufferLib}. For model-free RL, we used the PPO implementation from PureJaxRL \cite{lu2022discovered}, and implemented the environment in JAX \cite{jax2018github}. Finally, we ran each final circuit through a simple post-hoc optimization using simple circuit identities and gate commutation relations to reduce gate count and circuit depth, as detailed in \cref{app:post-hoc}.

\Cref{table_results} summarizes the lowest two-qubit gate count and corresponding circuit depths of the circuits found with each method compared to the previous state-of-the-art solution for each code. For the smallest codes, the state-of-the-art solutions come from the SAT method of Ref.~\cite{peham2024automated}, while for the 23-qubit Golay code we compare to the circuit from Ref.~\cite{webster2025} with 56 two-qubit gates and circuit depth 17 (or alternatively the circuit from Ref.~\cite{Paetznick2012} with 57 two-qubit gates and depth 7). 
For larger codes, we did not have state-of-the-art results to compare to, largely due to the fact that previous optimal methods such as that of Ref.~\cite{peham2024automated} cannot scale to codes of larger size. Therefore, as a baseline we took the original graphs coming from \textit{Stim} and applied LC-optimization via simulated annealing to reduce edge count~\footnote{In the LC-optimization, we only considered pivot operators which are a sequence of three local complementations $\LC_i\LC_j\LC_i$ for connected nodes $i,j$ since they preserve the bipartite property of the graphs in the case of CSS codes.}, as described in \cref{circuits}. Then, we reported the number of two-qubit gates and circuit depth as the number of edges and maximum degree of the optimized graph (corresponding to a circuit of $\CZ$ gates alone, as discussed in \cref{circuits}). We find that neither the Qiskit \cite{qiskit2024} SDK function \textit{GreedySynthesisClifford} nor the Qiskit transpilation tool at highest optimization level are able to optimize this $\CZ$ circuit.

We find that all of our methods match or improve significantly on the existing state of the art for all codes. For example, we match the two-qubit gate count of the optimal SAT circuits from Ref.~\cite{peham2024automated} for smaller codes and even manage to reduce depth in some cases. For the Golay code, both the beam search and model-free RL methods reduce the best known two-qubit gate count by $11$ gates ($\sim$20\%) while matching the state-of-the-art circuit depth, with \textit{QuSynth} being able to further reduce gate count by one. We also reduce the two-qubit gate count for preparation of the $144$-qubit Gross code by a factor of $2.5$ with respect to the result obtained from LC-optimization (or a factor of $5$ compared to the graph representative coming from \textit{Stim}). We remark that the circuits for larger codes were optimized for two-qubit gate count. It is likely that the circuit depth for these codes can be reduced, possibly at the cost of more two-qubit gates, by more aggressively enforcing depth constraints during graph decimation.

\begin{figure}
    \centering
    \includegraphics[width=\linewidth]{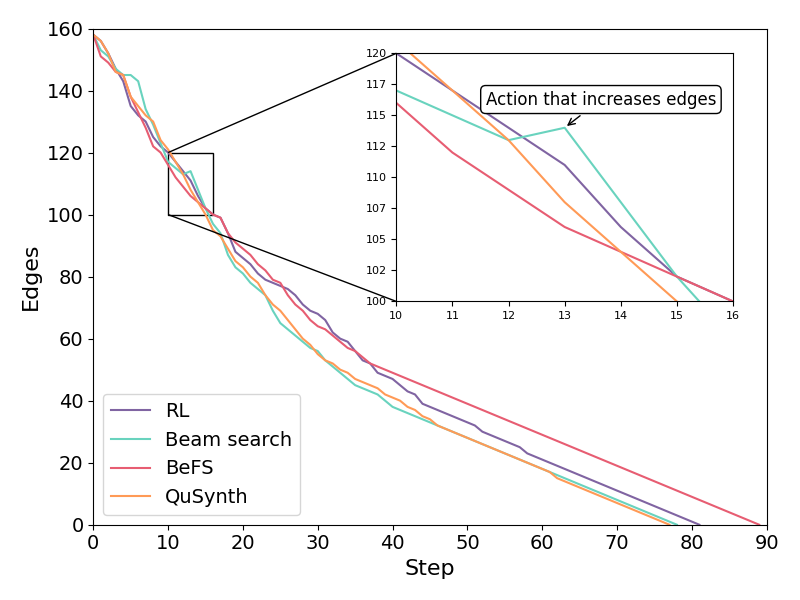}
    \caption{Remaining edges in the graph for the $\nkd{36}{8}{6}$ code: comparison between the best sequences obtained with BeFS, beam search, RL and \textit{QuSynth}, respectively. Beam search selects an action that increases the number of edges, but then reaches the zero-edge target state with fewer total steps than BeFS and RL. \textit{QuSynth}, instead, is able to find an overall better solution without ever increasing the number of edges.}
    \label{fig:36_edge_progression}
\end{figure}

When comparing between our different methods, we find that the more sophisticated methods are unable to improve upon BeFS for codes smaller than the Golay code (hence why they are not included in the table). However, for larger codes, the greedy nature of the BeFS algorithm leads to suboptimal results. This can be clearly seen by tracking the number of edges during graph decimation of the $\nkd{36}{8}{6}$ code when using beam search, as in \cref{fig:36_edge_progression}. There, we see that the number of edges in a more optimal solution actually increases at one step, but this allows the removal of more edges in later steps, eventually overtaking BeFS. We can also compare our reinforcement learning results against the heuristic search methods.
In almost all cases, our reinforcement learning methods exceed or match the heuristic search results, with
\textit{QuSynth} finding the best solution for almost every code. These methods achieve the overall best results by training on multi-step returns that enable credit assignment to actions over even longer horizons than beam search is able to consider. By 
using tree search, \textit{QuSynth} is able to find better solutions than the model-free reinforcement learning agent for intermediate-size codes.

We also experimented with using LC-optimized graphs as inputs to graph decimation (solutions indicated by $^\ddagger$ in \cref{table_results}). When using BeFS, we observed minimal change for small codes, but more significant effects for larger codes due to the large difference between the number of graph edges before and after optimization. Therefore, for the three largest codes, we also applied each of our methods to the LC-optimized graph, obtaining a considerable improvement in the reduction of gate count and depth. Using LC-optimization as a pre-processing step would also boost the ability to scale all the approaches to much larger codes, thanks to the reduction in the initial number of edges.

A key consideration for the choice of method is a tradeoff between the quality of the solution and the computational resources required. {Notably, the best solutions from BeFS for the codes smaller than the Golay code can be obtained in $\lesssim 10$ iterations, thereby giving state-of-the-art results in fractions of a second. However, for the Golay code and larger codes, BeFS requires many more iterations to obtain the best results, leading to much larger runtimes.} In \cref{sec:time_comparisons}, we compare the mean steps per second and the time to best solution for each of the optimization methods, obtained using an NVIDIA A100 80GB.
Though the time to solution is the main quantity of interest, we also record the number of environment steps per second to give a more comparable benchmark between methods with different best solutions.
Overall, beam search is the fastest to reach its best solution as reported in \cref{table_results}.
Reinforcement learning approaches take significantly longer than the other methods.
\textit{QuSynth} is substantially more computationally intensive than even the model-free agent, since it requires running a tree search for each move in self-play. As a result, it is the slowest method per iteration, but it is the most sample efficient as it requires a lower total number of iterations compared to beam search and RL, despite the longer total running times. 

\section{Discussion}
\label{sec:discussion}

In this section we discuss two aspects of the graph decimation approach and \textit{QuSynth}, namely applications beyond QEC state preparation, and routes to fault tolerance.

\subsection{Further applications of graph decimation}

We have focused on applying the graph decimation method to the preparation of code states of quantum error correcting codes. This is an important part of quantum error correction as many error correction primitives require the preparation of ancillary logical states (ideally in a fault-tolerant manner, see the next section). However, the graph decimation and \textit{QuSynth} can be applied more broadly. Here, we briefly describe a some other applications of graph decimation.

\paragraph{Measurement-based quantum computation.} One approach to universal quantum computation called measurement-based quantum computation (MBQC) involves the preparation of certain many-qubit resource states followed by adaptive single-qubit measurements \cite{Raussendorf2003measurementbased}. In the typical case, these resource states are stabilizer states, and graph decimation can naturally be applied to optimize their preparation, and thereby optimize MBQC as a whole. Even for simple resource states like graph states on square (triangular) lattices \cite{Raussendorf2001,VanDenNest2004}, one can derive state preparation circuits using $\CX$ ($\CY$) gates alongside $\CZ$ gates that reduce total TQ gate count by a factor of $\nicefrac{3}{4}$ $\left(\nicefrac{2}{3} \right)$ \footnote{For a square lattice, this can be accomplished by decomposing the edges of the graph into disjoint 4-cycles, one for every other plaquette, and then using the sequence of three gates shown in \cref{eq:4cycle} to prepare each 4-cycle in an appropriate order. A similar trick on the triangular lattice prepares each 3-cycle with one $\CZ$ and one $\CY$.}. When applying the method to more general resource states, especially algorithm-specific graph states in which Pauli measurements have been pre-computed (which tend to have highly connected graphs) \cite{vijayan2024compilation,kaldenbach2024mapping,kaldenbach2025efficientpreparationresourcestates,sunami2022graphixoptimizingsimulatingmeasurementbased,Ferguson2021}, we can find drastic savings in the number of TQ gates compared to circuits using $\CZ$ gates alone.

\paragraph{Hamiltonian simulation and variational circuits.} For many classes of quantum circuits, including trotterized Hamiltonian simulation and variational optimization circuits \cite{farhi2014quantum,tilly2022thevariational}, the goal is to implement a series of Pauli rotations, which are unitaries of the form $R_\theta(P) = e^{i\theta P}$ where $P$ is a potentially high-weight tensor product of Pauli operators on $N$ qubits.
In most architectures, particularly in a fault-tolerant context, these high-weight Pauli rotations need to be decomposed into Clifford gates and single-qubit Pauli rotations. That is, we find a Clifford circuit $C$ such that $ P =C Z_i C^\dagger $ for some qubit $i$, such that $R_\theta(P) = C R_\theta (Z_i) C^\dagger$. More generally, for $M \leq N$ commuting Paulis $P_i$, it is always possible to find a Clifford circuit $C$ such that $\prod_{i=1}^M R_{\theta_i}(P_i) = C \prod_{i=1}^M R_{\theta_i}(Z_i) C^\dagger$. Finding this Clifford can be broken into two steps \cite{GoubaultdeBrugiere2025graphstatebased}. We first find $C_1$ such that $C_1^\dagger P_i C_1$ is diagonal in the $Z$-basis, \textit{i.e.}, a tensor product of $I$ and $Z$ only, for each $i=1,\dots,M$. We then use standard gaussian elimination, or some optimized version thereof, to find a Clifford $C_2$ that simultaneously maps these $Z$-strings to distinct single-qubit $Z$ operators. The final Clifford is then $C = C_1 C_2$.

The first step, \textit{i.e.}, the problem of finding the Clifford circuit $C_1$ that simultaneously diagonalizes $N$ commuting Pauli operators of length $N$ is exactly the same problem as preparing the stabilizer state defined by those Pauli strings \footnote{Indeed, after graph decimation, our final state is the all-$\ket{+}$ product state. Thus, all initial stabilizers must have been mapped to stabilizers of this product state, which must be $X$-strings. These $X$-strings can be turned into $Z$ operators by application of a global Hadamard operation}. Therefore, we can directly apply graph decimation to optimize this step. More generally, we could consider a modified graph decimation algorithm that implements both steps simultaneously.

\paragraph{Clifford circuits and isometries.} Graph decimation works by expressing a stabilizer state in terms of a graph and sequentially removing edges from the graph using Clifford operations. Clifford operators, and more generally Clifford isometries (such as arbitrary encoding circuits for QEC codes), can also be expressed in terms of graphs \cite{Duncan2020graphtheoretic,GoubaultdeBrugiere2025graphstatebased}. However, unlike the representation of stabilizer states as graph states, not every edge in the graph representation of a Clifford isometry corresponds to a physical two-qubit gate. Rather, certain edges can only be removed using a sequence of gates coming from a gaussian elimination step. The general procedure of obtaining a circuit from a graph representation of a Clifford isometry is called circuit extraction \cite{Backens2021therebackagain}. Our graph decimation protocol can be easily modified to optimize circuit extraction, opening the way to optimization of general Clifford circuits and isometries.

\subsection{Fault tolerance}

The state preparation circuits that we generate are not fault-tolerant, meaning that a low-weight error arising from a faulty gate in the preparation circuit can spread to a high-weight error at the end of the circuit. Nevertheless, our circuits can be used as an initial step in the synthesis of fault-tolerant circuits; here we summarize three methods to accomplish this.

\paragraph{Verification circuits.} One approach is to follow the non-fault tolerant circuits derived from our method with a verification circuit that checks for high-weight errors \cite{Goto2016,peham2024automated}. These verification circuits are constructed by analyzing the various errors that can arise from each two-qubit gate fault and then identifying minimal sets of stabilizers that can be measured to catch such errors. Fault tolerance is then achieved by either post-selecting on successful checks or actively applying corrections \cite{heussen2023strategiesfor,schmid2025deterministicfaulttolerantstatepreparation}.

\paragraph{Flagging gadgets.} We can extend the idea of verification circuits by also checking for errors in the middle of the preparation circuit. This more general process is called flagging and involves using ancillary qubits to measure certain checks throughout the circuit, catching errors before the spread to something undetectable or uncorrectable~\cite{Chamberland2018flagfaulttolerant, chao2020flagFT, forlivesi2025}.

\paragraph{Entanglement purification.} We can also take several copies of the noisy states coming from our state preparation circuits and use entanglement purification to produce a single noiseless copy. In such a protocol, it is essential to avoid correlated errors between the different copies, since these would propagate through the distillation protocol. In Ref.~\cite{Paetznick2012}, this was achieved for the Golay code by using the permutation symmetries of the code to permute qubits of the final prepared state, resulting in different error models on each copy. More generally, one can try to find several inequivalent preparation circuits whose circuit-level errors result in distinct final errors on the prepared state \cite{weilandt2026synthesis}. Here, the many possible paths that can be taken during graph decimation could be used to design such inequivalent circuits while retaining low depth and TQ gate count.

\section{Conclusion} 
\label{sec:conclusion}

We have presented a general and scalable method for preparing stabilizer states with significantly reduced quantum resource requirements. By exploiting the graph state representation and formulating circuit synthesis as a sequential decision-making problem, we developed optimization techniques that reduce two-qubit gate counts by 40–80\% compared to $\CZ$-only circuits for a given initial unoptimized graph. Most notably, we demonstrate preparation of IBM's 144-qubit gross code using only 357 two-qubit gates, a $\sim 5 \times$ reduction with compared to the implementation with no optimized initial graph, representing the largest stabilizer state for which AI-based circuit synthesis has been successfully applied, extending well beyond the $\sim$20-qubit limit of prior methods \cite{zen2024quantum}.

Our approach achieves this scalability through a novel graph-based 
formulation combined with efficient implementations leveraging GPU parallelization. Solution times range from seconds for smaller codes to a few hours for the gross code, with beam search achieving over 60 million environment steps per second and model-free RL training exceeding 2 million steps per second.

These reductions in two-qubit gate count and circuit depth have immediate implications for near-term quantum computing, where two-qubit gates remain the dominant error source. By substantially reducing both metrics, our optimized circuits bring practical state preparation closer to current hardware capabilities, potentially enabling quantum error correction demonstrations that would otherwise be infeasible.

\acknowledgments
We thankfully acknowledge Marco Ballarin, Asmae Benhemou, Serban Cercelescu, Steve Clark, Ali Hussein, Pranav Kalidindi, Alex Koziell-Pipe, Selwyn Simsek and Basudha Srivastava for useful discussions. We are grateful to Stephen Clark and Oliver Hart for comments on the final manuscript.

\bibliography{bibliography}

\clearpage 
\appendix

\section{Graph State Decimation for CSS States}
\label{app:bipartite_graphs}
If a CSS state (whose stabilizer generators can be expressed as a set of all-$X$ operators and a set of all-$Z$ operators) is input, the graph to be decimated will be bipartite.
For these graphs, the greedy algorithm always outputs a circuit that can be recast into one consisting of $\CX$ gates alone, up to Hadamard gates at the start and end of the circuit.
To prove this, we note that, when the graph is bipartite, we can divide the nodes into two subsets $V=A\cup B$, such that all edges connect a node in $A$ to a node in $B$.
This has several implications.
Since edges only exist between $A$ and $B$, $\CZ$ gates will only ever be applied between $A$ and $B$, and $\CX$ gates will only act within $A$ or within $B$ (because this will be equivalent to a product of $\CZ$ gates between $A$ and $B$).
A $\CY$ gate will never be applied since it would combine the two effects, resulting in a $\CZ$ within $A$ or within $B$.
Therefore, if we conjugate the circuit that is output by the algorithm by Hadamard gates on all nodes in $A$, this will transform all $\CZ$ gates into $\CX$ gates with targets lying in $A$, and it will flip the control and target qubits of all $\CX$ gates acting within $A$.
The net result is a circuit consisting of only $\CX$ gates acting on initial $|0\rangle$ or $|+\rangle$ states followed by a single layer of Hadamard gates at the end.
This reasoning also holds for decimation algorithms that only apply gates which immediately reduce edge count.
For algorithms which do not always immediately reduce edge count, we observe that the output circuits still contain only $\ket{0}$ preparation, $\ket{+}$ preparation and \CX gates for all instances considered in this work.

\section{Size of action space for naive exhaustive algorithm}

\label{sec:combinatorial_space}

\begin{figure}
\centering 
\includegraphics[width=8.6cm]{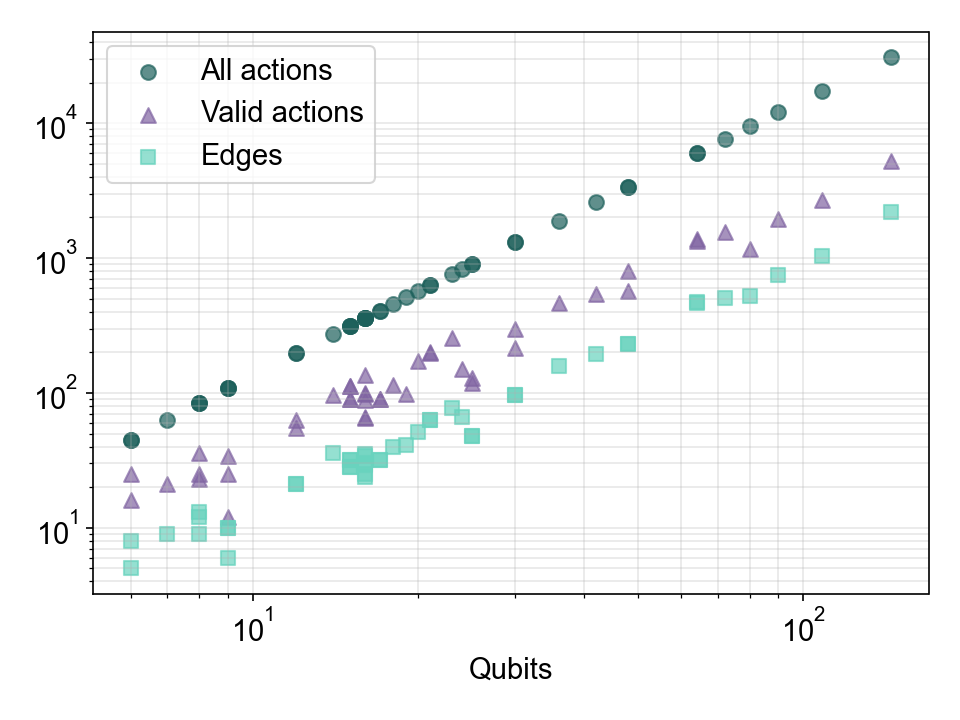}
\caption{Number of possible two-qubit gates (actions) and initial edge count for various stabilizer codes with differing physical qubit requirements. Valid actions are actions which remove at least one edge.} 
\label{fig:action_masking}
\end{figure}

In this section, we explicitly compute the number of states considered in a na\"{i}ve exhaustive algorithm that explores all actions and keeps all states at every step. Note that here we do not consider memoization of intermediate results. For $N_{Q}$ qubits, each possible $\CZ$ gate corresponds to an undirected qubit pair, of which there are $N_{Q}(N_{Q}-1)/2$. For $\CX$ and $\CY$ gates, the control and target qubits are distinct, resulting in $N_{Q}(N_{Q}-1)$ pairs for each gate type. The sum of possible $\CZ$, $\CX$, and $\CY$ gates gives the total number of actions available at a given step,
\begin{equation}
N_A^{\mathrm{tot}} = \frac{5}{2}N_{Q}(N_{Q}-1).
\label{eq:NA}
\end{equation}
For a sequence of $M$ actions, the state space can be visualized as a tree,
illustrated in \cref{fig:tree_of_states}, of depth $M$ with branching factor $N_A^{\mathrm{tot}}$ and a number of nodes equal to,
\begin{equation}
N_{states} = \sum_{i=1}^{M} (N_A^{\mathrm{tot}})^i = \frac{(N_A^{\mathrm{tot}})^{M+1} - N_A^{\mathrm{tot}}}{N_A^{\mathrm{tot}} - 1}.
\label{eq:NS}
\end{equation}
From \eqref{eq:NA} and \eqref{eq:NS}, the total combinations of possible circuits scales as $N_Q^{2M}$, which precludes exhaustive search and requires efficient search and other optimization techniques to find improved solutions. 
We note that the size of the action space is substantially reduced by considering only actions that remove at least one edge. This is shown in \cref{fig:action_masking}.

\section{Post-hoc compilation and depth optimization}
\label{app:post-hoc}
\begin{figure}[H]
\centering
\begin{tikzpicture}
\node at (-3.2, 2.8) {a)};
\node at (-1.75,2) {\begin{tikzpicture}[scale=1.000000,x=1pt,y=1pt]
\filldraw[color=white] (0.000000, -7.500000) rectangle (54.000000, 37.500000);
\draw[color=black] (0.000000,30.000000) -- (54.000000,30.000000);
\draw[color=black] (0.000000,30.000000) node[left] {$a$};
\draw[color=black] (0.000000,15.000000) -- (54.000000,15.000000);
\draw[color=black] (0.000000,15.000000) node[left] {$b$};
\draw[color=black] (0.000000,0.000000) -- (54.000000,0.000000);
\draw[color=black] (0.000000,0.000000) node[left] {$c$};
\draw (9.000000,30.000000) -- (9.000000,15.000000);
\begin{scope}
\draw[fill=white] (9.000000, 15.000000) circle(3.000000pt);
\clip (9.000000, 15.000000) circle(3.000000pt);
\draw (6.000000, 15.000000) -- (12.000000, 15.000000);
\draw (9.000000, 12.000000) -- (9.000000, 18.000000);
\end{scope}
\filldraw (9.000000, 30.000000) circle(1.500000pt);
\draw (27.000000,15.000000) -- (27.000000,0.000000);
\filldraw (27.000000, 15.000000) circle(1.500000pt);
\filldraw (27.000000, 0.000000) circle(1.500000pt);
\draw (45.000000,30.000000) -- (45.000000,15.000000);
\begin{scope}
\draw[fill=white] (45.000000, 15.000000) circle(3.000000pt);
\clip (45.000000, 15.000000) circle(3.000000pt);
\draw (42.000000, 15.000000) -- (48.000000, 15.000000);
\draw (45.000000, 12.000000) -- (45.000000, 18.000000);
\end{scope}
\filldraw (45.000000, 30.000000) circle(1.500000pt);
\end{tikzpicture}};
\node at (0,2) {$\mapsto$};
\node at (1.75,2) {\begin{tikzpicture}[scale=1.000000,x=1pt,y=1pt]
\filldraw[color=white] (0.000000, -7.500000) rectangle (36.000000, 37.500000);
\draw[color=black] (0.000000,30.000000) -- (36.000000,30.000000);
\draw[color=black] (0.000000,30.000000) node[left] {$a$};
\draw[color=black] (0.000000,15.000000) -- (36.000000,15.000000);
\draw[color=black] (0.000000,15.000000) node[left] {$b$};
\draw[color=black] (0.000000,0.000000) -- (36.000000,0.000000);
\draw[color=black] (0.000000,0.000000) node[left] {$c$};
\draw (9.000000,30.000000) -- (9.000000,0.000000);
\filldraw (9.000000, 30.000000) circle(1.500000pt);
\filldraw (9.000000, 0.000000) circle(1.500000pt);
\draw (27.000000,15.000000) -- (27.000000,0.000000);
\filldraw (27.000000, 15.000000) circle(1.500000pt);
\filldraw (27.000000, 0.000000) circle(1.500000pt);
\end{tikzpicture}};
\node at (-3.2, 0.8) {b)};
\node at (-1.75,0) {\begin{tikzpicture}[scale=1.000000,x=1pt,y=1pt]
\filldraw[color=white] (0.000000, -7.500000) rectangle (54.000000, 37.500000);
\draw[color=black] (0.000000,30.000000) -- (54.000000,30.000000);
\draw[color=black] (0.000000,30.000000) node[left] {$a$};
\draw[color=black] (0.000000,15.000000) -- (54.000000,15.000000);
\draw[color=black] (0.000000,15.000000) node[left] {$b$};
\draw[color=black] (0.000000,0.000000) -- (54.000000,0.000000);
\draw[color=black] (0.000000,0.000000) node[left] {$c$};
\draw (9.000000,30.000000) -- (9.000000,15.000000);
\begin{scope}
\draw[fill=white] (9.000000, 15.000000) circle(3.000000pt);
\clip (9.000000, 15.000000) circle(3.000000pt);
\draw (6.000000, 15.000000) -- (12.000000, 15.000000);
\draw (9.000000, 12.000000) -- (9.000000, 18.000000);
\end{scope}
\filldraw (9.000000, 30.000000) circle(1.500000pt);
\draw (27.000000,30.000000) -- (27.000000,0.000000);
\filldraw (27.000000, 30.000000) circle(1.500000pt);
\filldraw (27.000000, 0.000000) circle(1.500000pt);
\draw (45.000000,15.000000) -- (45.000000,0.000000);
\filldraw (45.000000, 15.000000) circle(1.500000pt);
\filldraw (45.000000, 0.000000) circle(1.500000pt);
\end{tikzpicture}};
\node at (0,0) {$\mapsto$};
\node at (1.75,0) {\begin{tikzpicture}[scale=1.000000,x=1pt,y=1pt]
\filldraw[color=white] (0.000000, -7.500000) rectangle (36.000000, 37.500000);
\draw[color=black] (0.000000,30.000000) -- (36.000000,30.000000);
\draw[color=black] (0.000000,30.000000) node[left] {$a$};
\draw[color=black] (0.000000,15.000000) -- (36.000000,15.000000);
\draw[color=black] (0.000000,15.000000) node[left] {$b$};
\draw[color=black] (0.000000,0.000000) -- (36.000000,0.000000);
\draw[color=black] (0.000000,0.000000) node[left] {$c$};
\draw (9.000000,15.000000) -- (9.000000,0.000000);
\filldraw (9.000000, 15.000000) circle(1.500000pt);
\filldraw (9.000000, 0.000000) circle(1.500000pt);
\draw (27.000000,30.000000) -- (27.000000,15.000000);
\begin{scope}
\draw[fill=white] (27.000000, 15.000000) circle(3.000000pt);
\clip (27.000000, 15.000000) circle(3.000000pt);
\draw (24.000000, 15.000000) -- (30.000000, 15.000000);
\draw (27.000000, 12.000000) -- (27.000000, 18.000000);
\end{scope}
\filldraw (27.000000, 30.000000) circle(1.500000pt);
\end{tikzpicture}};
\node at (-3.2, -1.2) {c)};
\node at (-1.75,-2) {\begin{tikzpicture}[scale=1.000000,x=1pt,y=1pt]
\filldraw[color=white] (0.000000, -7.500000) rectangle (54.000000, 37.500000);
\draw[color=black] (0.000000,30.000000) -- (54.000000,30.000000);
\draw[color=black] (0.000000,30.000000) node[left] {$a$};
\draw[color=black] (0.000000,15.000000) -- (54.000000,15.000000);
\draw[color=black] (0.000000,15.000000) node[left] {$b$};
\draw[color=black] (0.000000,0.000000) -- (54.000000,0.000000);
\draw[color=black] (0.000000,0.000000) node[left] {$c$};
\draw (9.000000,30.000000) -- (9.000000,15.000000);
\begin{scope}
\draw[fill=white] (9.000000, 15.000000) circle(3.000000pt);
\clip (9.000000, 15.000000) circle(3.000000pt);
\draw (6.000000, 15.000000) -- (12.000000, 15.000000);
\draw (9.000000, 12.000000) -- (9.000000, 18.000000);
\end{scope}
\filldraw (9.000000, 30.000000) circle(1.500000pt);
\draw (27.000000,15.000000) -- (27.000000,0.000000);
\begin{scope}
\draw[fill=white] (27.000000, 0.000000) circle(3.000000pt);
\clip (27.000000, 0.000000) circle(3.000000pt);
\draw (24.000000, 0.000000) -- (30.000000, 0.000000);
\draw (27.000000, -3.000000) -- (27.000000, 3.000000);
\end{scope}
\filldraw (27.000000, 15.000000) circle(1.500000pt);
\draw (45.000000,30.000000) -- (45.000000,15.000000);
\begin{scope}
\draw[fill=white] (45.000000, 15.000000) circle(3.000000pt);
\clip (45.000000, 15.000000) circle(3.000000pt);
\draw (42.000000, 15.000000) -- (48.000000, 15.000000);
\draw (45.000000, 12.000000) -- (45.000000, 18.000000);
\end{scope}
\filldraw (45.000000, 30.000000) circle(1.500000pt);
\end{tikzpicture}};
\node at (0,-2) {$\mapsto$};
\node at (1.75,-2) {\begin{tikzpicture}[scale=1.000000,x=1pt,y=1pt]
\filldraw[color=white] (0.000000, -7.500000) rectangle (36.000000, 37.500000);
\draw[color=black] (0.000000,30.000000) -- (36.000000,30.000000);
\draw[color=black] (0.000000,30.000000) node[left] {$a$};
\draw[color=black] (0.000000,15.000000) -- (36.000000,15.000000);
\draw[color=black] (0.000000,15.000000) node[left] {$b$};
\draw[color=black] (0.000000,0.000000) -- (36.000000,0.000000);
\draw[color=black] (0.000000,0.000000) node[left] {$c$};
\draw (9.000000,30.000000) -- (9.000000,0.000000);
\begin{scope}
\draw[fill=white] (9.000000, 0.000000) circle(3.000000pt);
\clip (9.000000, 0.000000) circle(3.000000pt);
\draw (6.000000, 0.000000) -- (12.000000, 0.000000);
\draw (9.000000, -3.000000) -- (9.000000, 3.000000);
\end{scope}
\filldraw (9.000000, 30.000000) circle(1.500000pt);
\draw (27.000000,15.000000) -- (27.000000,0.000000);
\begin{scope}
\draw[fill=white] (27.000000, 0.000000) circle(3.000000pt);
\clip (27.000000, 0.000000) circle(3.000000pt);
\draw (24.000000, 0.000000) -- (30.000000, 0.000000);
\draw (27.000000, -3.000000) -- (27.000000, 3.000000);
\end{scope}
\filldraw (27.000000, 15.000000) circle(1.500000pt);
\end{tikzpicture}};
\end{tikzpicture}
\caption{Post-hoc circuit simplification rules applied to circuits output by graph-search algorithms used in this work. Time-reversed versions of these rules are also applied.}
\label{fig:rules}
\end{figure}
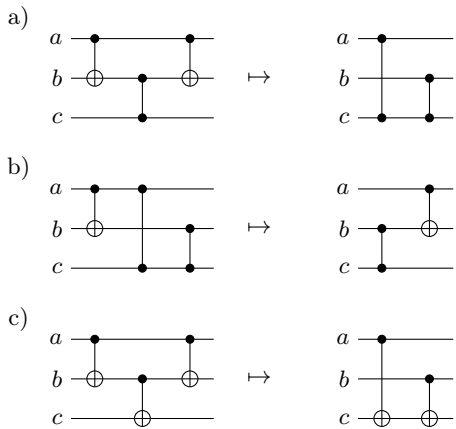

In order to improve the optimization, we run a post-hoc compilation step on each circuit which looks for specific gates sequences, verifies commutation relations and replaces them with shorter ones, reducing the gate count of the circuit, see \cref{fig:rules}.

Additionally, at the end of the circuit preparation, we perform a depth optimization pass, which greedily reorders gates to minimize circuit depth while respecting commutativity constraints. The algorithm processes gates sequentially, attempting to place each gate in the earliest valid layer by scanning backwards from the current final layer. A gate can be placed in an earlier layer if it acts on disjoint qubits from all gates in that layer. When a gate shares qubits with gates in a layer but commutes with them, the algorithm continues searching earlier layers; otherwise, it terminates the search and places the gate in the earliest valid position found.

\section{Solution time comparisons}
\label{sec:time_comparisons}

\begin{figure} [ht!]
\centering
\includegraphics[width=8.6cm]{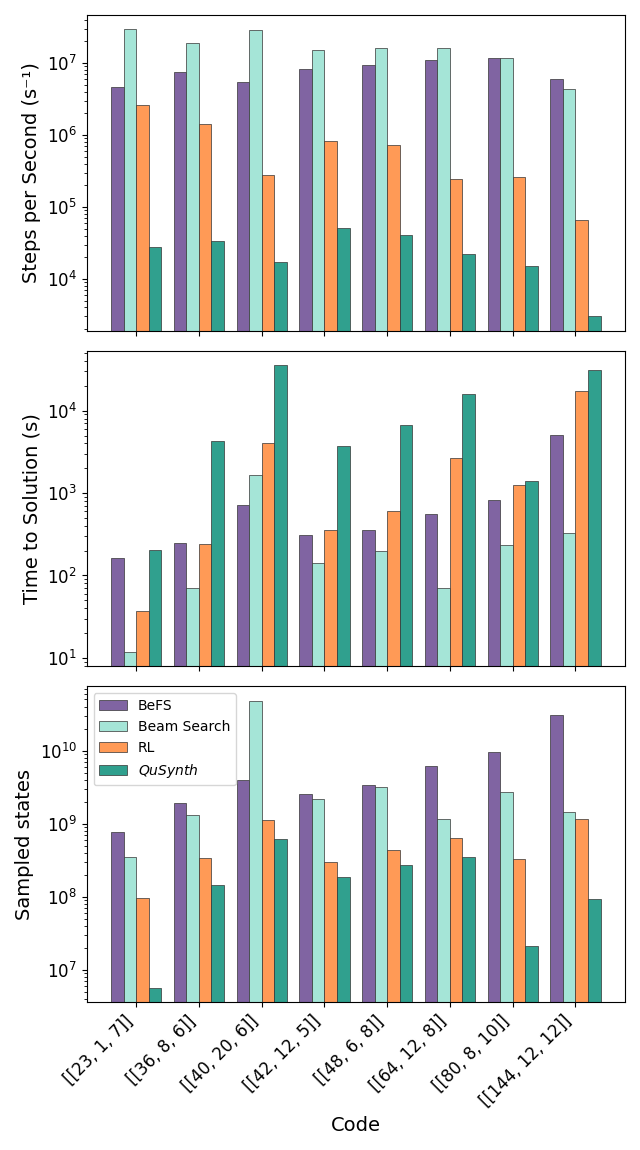}
\caption{Comparison of environment steps per second (top), time to best solution (center) and number of sampled states (bottom, lower means higher sample-to-solution efficiency), for solution methods on selected stabilizer codes. All results obtained using one Nvidia A100 80GB. The best-first search (BeFS) results are shown for a fixed number of 1M iterations and with the tie-breaker leading to the best solution (as specified in \cref{table_results}). \textit{QuSynth} is the only method not implemented in JAX and not using just-in-time compilation: as a result, despite being the slowest, it's the most efficient due to the low sample rate. Sample-to-solution efficiency is inversely proportional to the total number of states visited to get the best solution of each method, which is in turn given by the product of the steps rate and the time to solution.}
\label{fig:times}
\end{figure}

In \cref{fig:times}, we compare the mean steps per second and the time to best solution for each of the optimization methods, obtained using a NVIDIA A100 80GB at maximum memory capacity. 

Environment steps per second is a commonly used metric in RL \cite{petrenko2020samplefactoryegocentric3d} to assess the rate at which data can be acquired and trained on or, in the case of BeFS or beam search, evaluated for a solution.
\Cref{fig:times} (upper plot) shows the steps per second (SPS) that each method can achieve, including all other algorithmic overhead such as updating circuits, logging metrics, and updating model weights. For RL, we use the hyperparameters given in \cref{sec:hyperparams} but increase the parallelization of environments to the maximum attainable within 80GB of memory, to show the maximum throughput possible on the device. For BeFS, the SPS increases from 3M to 6M as the code size increases because the number of possible actions per state increases and are computed in parallel. The SPS dips for the 144-qubit gross code, likely due to increased overhead of memory access to update larger adjacency matrices. For beam search, the SPS peaks at over 60M and decreases with increasing code size to 4M for the gross code, as the beam width must reduce to fit the larger matrices in memory. Using action masking decreases SPS for beam search by approximately 50\% for smaller codes and relatively less for larger codes. 

SPS for RL decreases with increased code size but achieves training at over 2M SPS for Golay code and 60k SPS for the gross code. 1M SPS training for RL, depending on the environment, is considered high throughput \cite{suarez2025pufferlib}, therefore our implementation is efficient. The RL implementation is at a slight disadvantage compared to the search methods, as it requires 32-bit data types for stable training, whereas the search methods use 8 and 16-bit datatypes to permit approximately 2x faster computation on GPU and lower memory requirements.

The lower plot of \cref{fig:times} shows the time to solution for each method per code, for the solutions in \cref{table_results}. BeFS is fastest, returning a Golay solution in less than one second and gross code in less than 100. Beam search is slightly slower, with much better results, and can complete within 300s for the gross code. RL takes significantly longer than the other methods, using the hyperparameters listed in \cref{sec:hyperparams}. The gross code solution, while substantially better than that found by beam search or BeFS, required almost 5 hours of training time, not including the time spent on hyperparameter tuning runs. Nonetheless, 5 hours is an acceptable amount of time to spend on computing a preparation circuit that will be used many times for state preparation, and all methods can leverage further parallelization and continual computational hardware improvements to scale to larger graph states in future.

\section{Other Reinforcement Learning Approaches}
\label{app:other_optim}

\subsection{Policy-guided beam search (PGBS)} 
\label{sec:PGBS}
In contrast to the default evaluation of a trained RL agent, which provides one solution, we combine the beam search exploration algorithm with the learned probability distribution over the action space of the policy network of a trained RL model. The so-called policy-guided beam search (PGBS) \cite{Huber2022, CHEN2025104351} replaces the random sampling of actions of the beam search with the sampling based on the model's policy distribution. This results in higher action sampling efficiency than using the beam search itself, allowing at the same time for more exploration than the default evaluation of the RL agent. On the one hand, this approach reduces back to the default evaluation in the limit of beam width of $N_A = 1$ actions and a cut-off of $N_S = 1$ states at each iteration. On the other hand, when a well-trained RL model with low entropy in the action probability distribution is deployed, only the single best action according to the learned policy is sampled at each step, obtaining no better result than the default evaluation. Overall, PGBS allows to reach better results when the RL model is not optimally trained, 
and reduces the RL results for $\nkd{36}{8}{6}$ from 81 to 78.

\section{Behavior Cloning with Transformers}
\label{sec:transformer_bc}

A shortcoming of MLPs is they are unable to generalize to unseen codes because their input size is specific to the code's qubit count, and training on new problem distributions has been shown to cause MLPs to ``forget'' previous learning  \cite{MCCLOSKEY1989109}. The Transformer architecture can overcome the limitations of MLPs by accepting variable-length input sequences through its self-attention mechanism \cite{NIPS2017_Attention}, enabling a single model to be trained simultaneously on circuits from codes of different sizes. This eliminates the need for sequential training on different problem distributions, thereby avoiding the catastrophic forgetting that affects MLPs when trained sequentially on new data distributions. Transformers also have the expressivity of GNNs and are able to capture complex relationships between different parts of the input without suffering from topology-related information bottlenecks \cite{joshi2025transformersgraphneuralnetworks}. 

Transformers are known to be unstable when trained in an RL setting, owing the sensitivity of the architecture to the non-stationary data distribution in RL tasks \cite{parisotto2019stabilizingtransformersreinforcementlearning}. Enabling the advantages of the Transformer architecture to be applied to the graph state synthesis task therefore requires a supervised learning formulation. Behavior cloning \cite{ross2011reduction, daftry2016learning} is a well-known framework to learn policies from expert demonstrations. We therefore use a behavior cloning approach, with trained MLP policies as experts, to train a single Transformer model that can synthesize state preparation circuits for unseen graph states.

Our approach is to assemble a dataset of observations and action distirubtions from each of the expert MLPs for the selected stabilizer codes in \cref{table_results}. In addition to the expert demonstrations, we increase the dataset size by using beam search to find optimized circuits for an additional 5 graph states, giving 13 total examples for stabilizer codes in the 23 to 144 qubit range. We use the best circuits from beam search runs to construct one-hot action encodings as the action distributions for the dataset. The observations in the dataset are the adjacency matrices of the states, concatenated with the eigenvectors of the graph Laplacian, which serve as input sequences to the Transformer. Each $i^{th}$ element of the input sequence to the Transformer is therefore a vector of dimension $2N_Q$, where $N_Q$ is number of qubits, with $N_Q$ elements from the $i^{\textrm{th}}$ row of the adjacency matrix and $N_Q$ as the $i^{\textrm{th}}$ elements from the eigenvectors of the graph Laplacian.  We find that including eigenvectors of the graph Laplacian are essential for the Transformer to learn, which has been observed in other works on Transformers for graph learning \cite{ying2021transformersreallyperformbad}, and serve as a form of positional encoding for a node within the graph structure. A cross-entropy loss function is used such that the output logits of the Transformer are trained to match the expert distributions.

Our experiments found that we could successfully clone the behaviors of all demonstrations into a single Transformer model but it would not generalize to unseen codes. By training on 12/13 experts demonstrations, training loss would decrease to zero and the target circuits could be synthesized by the model, however the circuits for the remaining held-out code, used for evaluation, were equivalent to random guessing. Dataset augmentation, through permutation of node labels and corresponding action distributions, did not improve the evaluation. We conclude that further research is required to improve the generalization performance, perhaps through increased dataset size with generated graphs. We report these findings here as a first step in this direction and suggest it as an area for further research.

\newpage

\begin{widetext}

\section{Hyperparameters and training dynamics} 
\label{sec:hyperparams}

\begin{table*}[htbp]
\centering
\small
\setlength{\tabcolsep}{4pt}
\renewcommand{\arraystretch}{1.2}
\definecolor{tealrow}{HTML}{E1F6F2}
\begin{tabular}{l c c c c c c c c}
\toprule
\textbf{Code} & $\nkd{23}{1}{7}$ & $\nkd{36}{8}{6}$ & $\nkd{40}{20}{6}$ & $\nkd{42}{12}{5}$ & $\nkd{48}{6}{8}$ & $\nkd{64}{12}{8}$ & $\nkd{80}{8}{10}$ & $\nkd{144}{12}{12}$ \\
\midrule
\multicolumn{9}{c}{Timesteps and parallelization} \\
\midrule
\rowcolor{tealrow} Total timesteps & 2.2M & 200M & 1.1B & 200M & 1.2B & 1.2B & 340M & 1B \\
Steps per logging increment & 96k & 2M & 4M & 2M & 20M & 4M & 4M & 4M \\
\rowcolor{tealrow} Number of environments & 100 & 4096 & 2048 & 1024 & 4096 & 2048 & 920 & 110 \\
\midrule
\multicolumn{9}{c}{PPO parameters} \\
\midrule
Clip $\epsilon$ & 0.14 & 0.16 & 0.15 & 0.15 & 0.15 & 0.15 & 0.15 & 0.15 \\
\rowcolor{tealrow}Discount $\gamma$ & 0.953 & 0.999 & 0.999 & 0.999 & 0.9999 & 0.999 & 0.999 & 0.9999 \\
GAE $\lambda$ & 0.995 & 0.995 & 0.998 & 0.99 & 0.999 & 0.998 & 0.998 & 0.999 \\
\rowcolor{tealrow} Rollout length & 64 & 128 & 256 & 160 & 164 & 256 & 128 & 512 \\
Entropy schedule & constant & cosine & cosine & cosine & linear & cosine & cosine & cosine \\
\rowcolor{tealrow} Entropy coefficient & 0.003 & 0.02 & 0.06 & 0.01 & 0015 & 0.06 & 0.06 & 0.04 \\
Entropy end fraction & 1.0 & 0.1 & 0.1 & 0.0001 & 0.0001 & 0.1 & 0.1 & 0.1 \\
\rowcolor{tealrow} Value function coefficient & 0.44 & 0.005 & 0.004 &0.01 & 0.03 & 0.004 & 0.0001 & 0.006 \\
\midrule
\multicolumn{9}{c}{Training parameters} \\
\midrule
Learning rate & 0.002 & 0.01 & 0.001 & 0.0015 & 0.0005 & 0.001 & 0.001 & 0.001 \\
\rowcolor{tealrow} Learning rate schedule & cosine & cosine & cosine & cosine & cosine & cosine & constant & cosine \\
LR end fraction & 0.28 & 0.05 & 0.5 & 0.1 & 0.1 & 0.5 & 0.5 & 0.3 \\
\rowcolor{tealrow} Update epochs & 2 & 3 & 3 & 3 & 1 & 3 & 3 & 3 \\
Number of mini-batches & 2 & 1 & 1 & 1 & 1 & 1 & 1 & 1 \\
\midrule
\multicolumn{9}{c}{Neural network parameters} \\
\midrule
\rowcolor{tealrow} Number of layers & 2 & 2 & 2 & 2 & 2 & 2 & 2 & 2 \\
Number of units & 64 & 64 & 256 & 160 & 256 & 256 & 256 & 512 \\
\bottomrule
\end{tabular}
\caption{Hyperparameters used for training the RL agent for each stabilizer code. Standard metric prefix notation is adopted, with k = $10^3$, M = $10^6$, B = $10^9$.}
\label{table:hyperparam_results}
\end{table*}

\end{widetext}

\subsection{Beam search parameters}
\label{sec:beam_search_params}

In contrast to BeFS, beam search has two additional parameters: beam width ($N_S$) and actions at each step ($N_A$). To investigate the effect of these parameters on the resulting circuits, we conducted a grid search of parameter combinations: $N_S$ from 10 to 10000 states and $N_A$ from 2\% to 100\% of possible actions. For each parameter combination, we ran beam search on the $\nkd{23}{1}{7}$ Golay code for 1000 iterations, with and without action masking of 759 possible actions. 

\begin{figure*}[ht]
\centering
\includegraphics[width=0.9\linewidth]{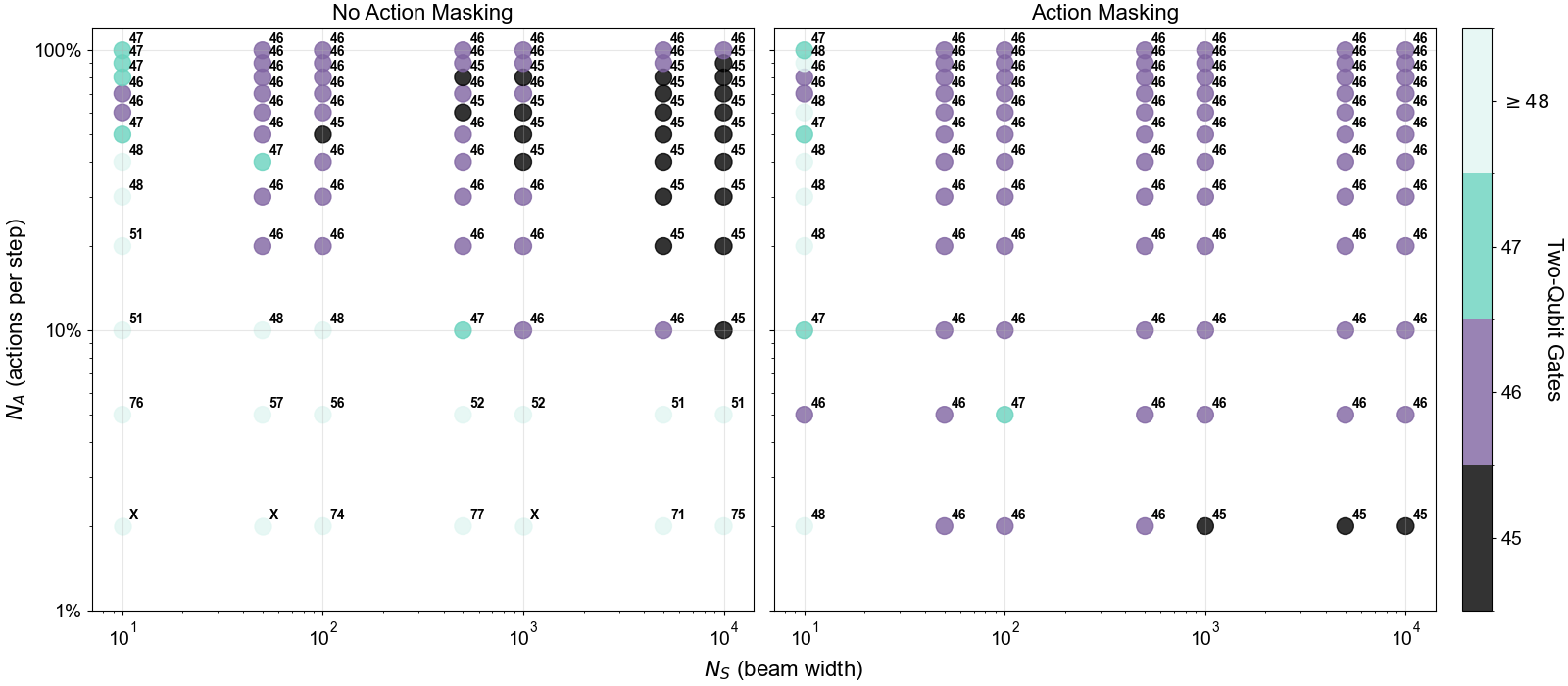}
\caption{Two-qubit gate count for combinations of beam search parameters $N_S$ and $N_A$ for the $\nkd{23}{1}{7}$ Golay code, without (left) and with (right) invalid action masking. `x' indicates no solution found.}
\label{fig:beam_search_params} 
\end{figure*}

\Cref{fig:beam_search_params} shows the minimum two-qubit gate counts resulting from these parameter combinations, with 45 gates as the best solution for the beam search. Without action masking, the algorithm reliably finds the 45-gate solution at larger beam widths and using 10-90\% of total actions. Notably, using 100\% of actions at each iteration, as BeFS does, never finds the 45-gate solution with beam search. We suggest this is because some locally suboptimal but globally optimal actions are necessary to reach the 45-gate solution and allowing some stochasticity in the actions applied at each step aids exploration. With action masking, $N_A$ of only 2\% can find the 45-gate solution with sufficient beam width. For beam search, action masking prioritizes actions that remove at least one edge, but other ``invalid'' actions can still be selected. This is what enables beam search to find a better solution to the $\nkd{36}{8}{6}$ code (see \cref{table_results}) than masked RL. We used this investigation to inform our choice of parameters for all codes, preferring to use action masking with $N_A$ set to a small percentage ($<5\%$) of total actions and beam width as large as possible within memory.

\subsection{RL training dynamics}
\label{sec:rl_training}

\begin{figure}
\centering
\includegraphics[width=8.6cm]{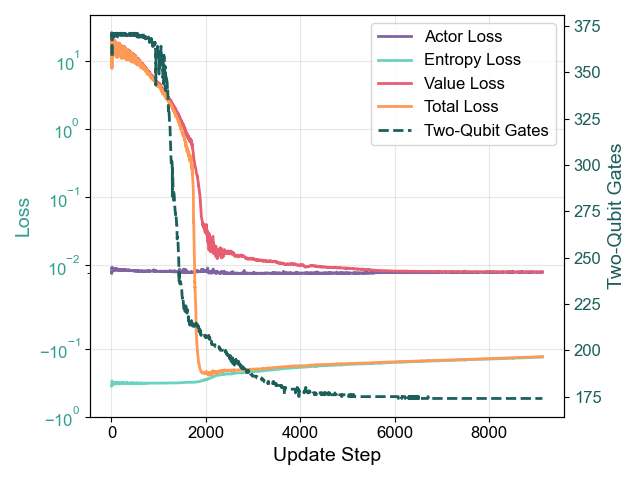}
\caption{RL agent training loss components and progression of two-qubit gate count for $\nkd{64}{12}{8}$ code.} 
\label{fig:training_loss}
\end{figure}

The progress of the two-qubit gate count during RL training followed a similar pattern for all codes, exemplified in the training curve of the $\nkd{64}{12}{8}$ code in \cref{fig:training_loss} (dark green). The value of the total loss and loss components (\cref{eq:ppo_loss}) in \cref{fig:training_loss} illustrate the characteristic phases of PPO training. Initially, the total loss (orange) is dominated by the value loss (red), which is high because the critic network has not yet learned to predict episode returns accurately. As the critic improves, the value loss decreases sharply, coinciding with the onset of the rapid decrease in gate count around update step 1500. This correlation suggests that accurate value estimation is a prerequisite for effective policy improvement.

The entropy term (cyan) decreases in magnitude, indicating that the policy distribution becomes more deterministic. The total loss becomes negative once dominated by the entropy term, and subsequently increases slightly as the policy converges, while value loss decreases towards zero. The relative contributions of these loss components were tuned through their coefficients to improve RL performance on this task, with values shown in \cref{sec:hyperparams}.

\section{Example quantum circuits}
We present here the circuits with lowest two-qubit gate count found by our optimization methods for the smallest selected stabilizer codes. 
Note that, for each of these circuits, if the ending Hadamard gates are pulled through to the beginning of the circuit, all two-qubit gates become $\CX$ gates.

\begin{figure}[H]
\centering
\includegraphics[width=0.99\linewidth]{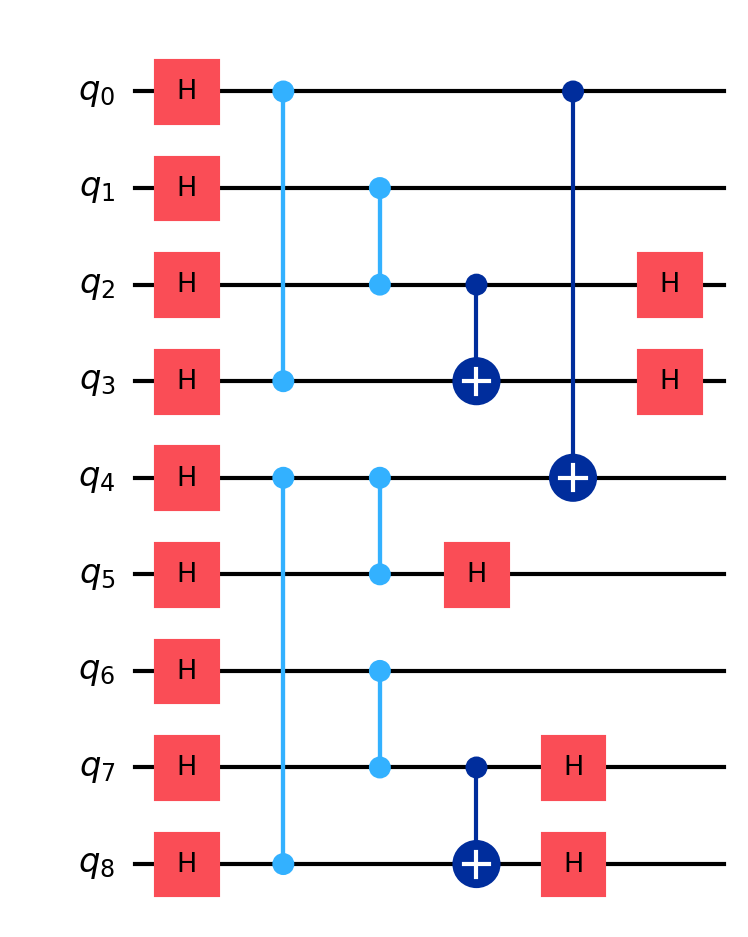}
\caption{Quantum circuit for preparing the $\ket{0}_L$ state of the $\nkd{9}{1}{3}$ surface code, obtained with BeFS. The red single-qubit gates are Hadamard; the light blue and dark blue two-qubit gates are the $\CZ$ and $\CX$ entangling gates, respectively.}
\label{fig:}
\end{figure}

\begin{figure}[H]
\centering
\includegraphics[width=0.99\linewidth]{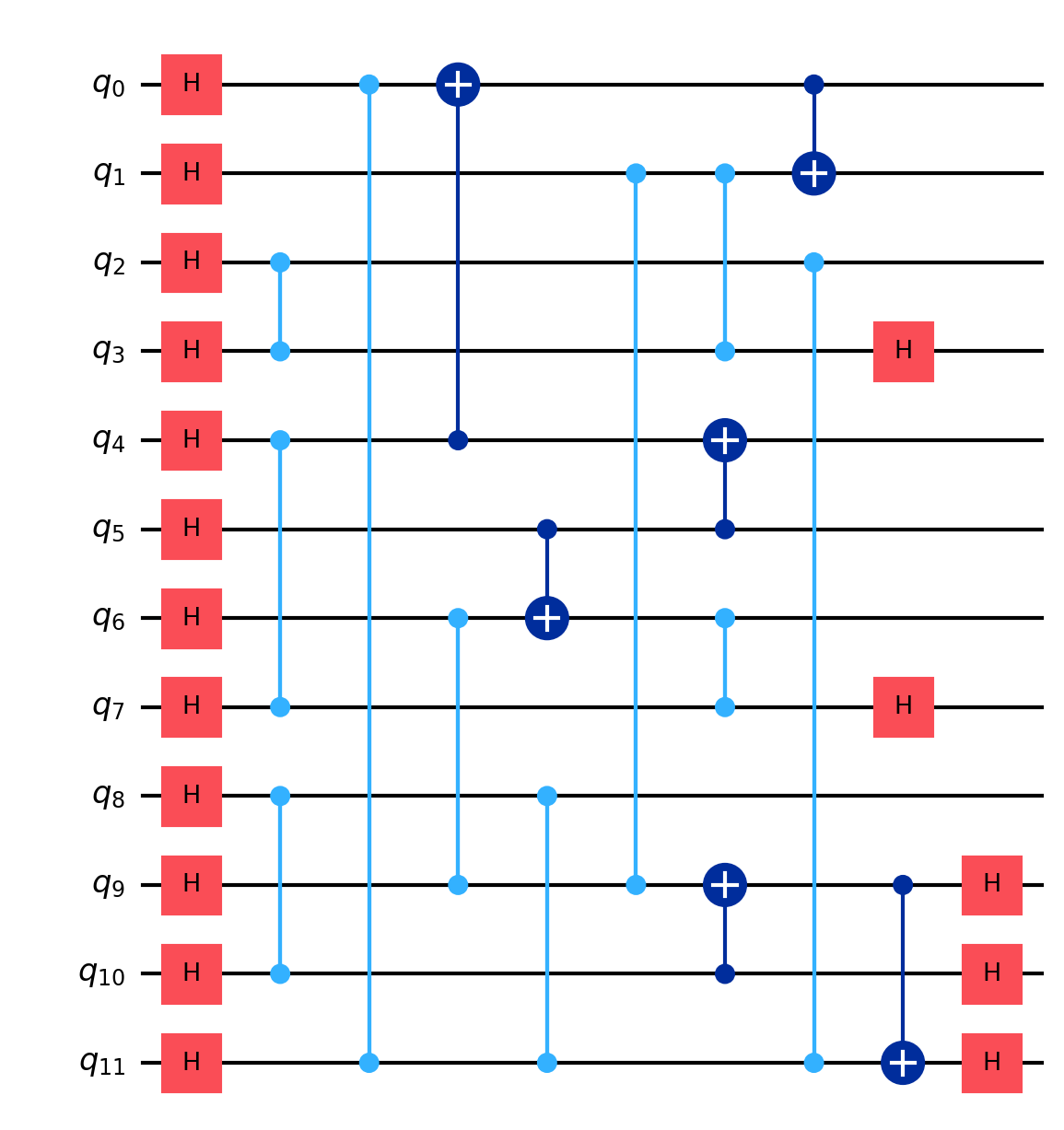}
\caption{Quantum circuit for preparing the $\ket{0}_L$ state of the $\nkd{12}{2}{4}$ Carbon code, obtained with BeFS. The red single-qubit gates are Hadamard; the light blue and dark blue two-qubit gates are the $\CZ$ and $\CX$ entangling gates, respectively.}
\label{fig:}
\end{figure}

\begin{figure}[H]
\centering
\includegraphics[width=0.99\linewidth]{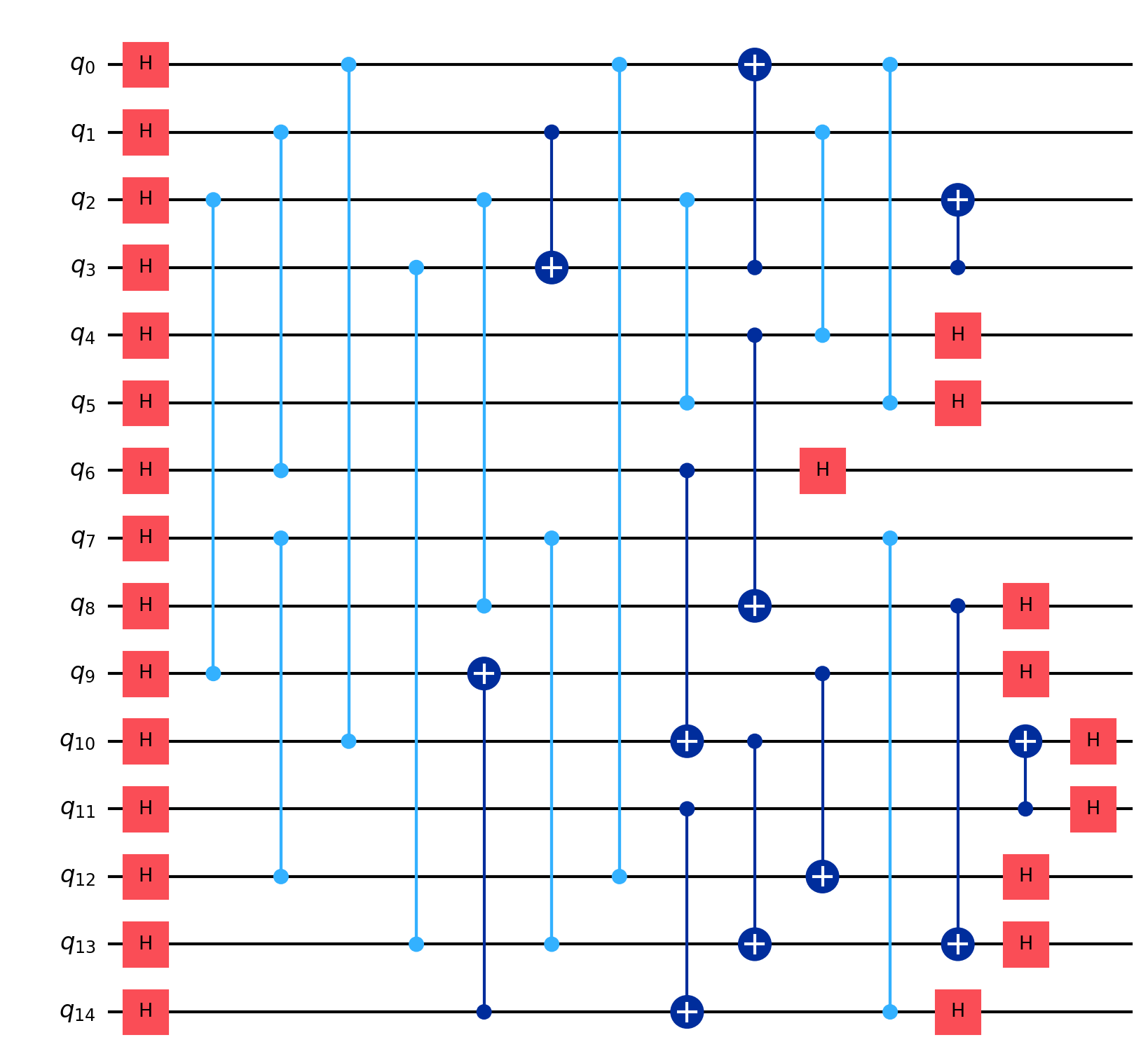}
\caption{Quantum circuit for preparing the $\ket{+}_L$ state of the $\nkd{15}{1}{3}$ Reed-Muller code, obtained with BeFS. The red single-qubit gates are Hadamard; the light blue and dark blue two-qubit gates are the $\CZ$ and $\CX$ entangling gates, respectively.}
\label{fig:}
\end{figure}

\begin{figure}[H]
\centering
\includegraphics[width=0.99\linewidth]{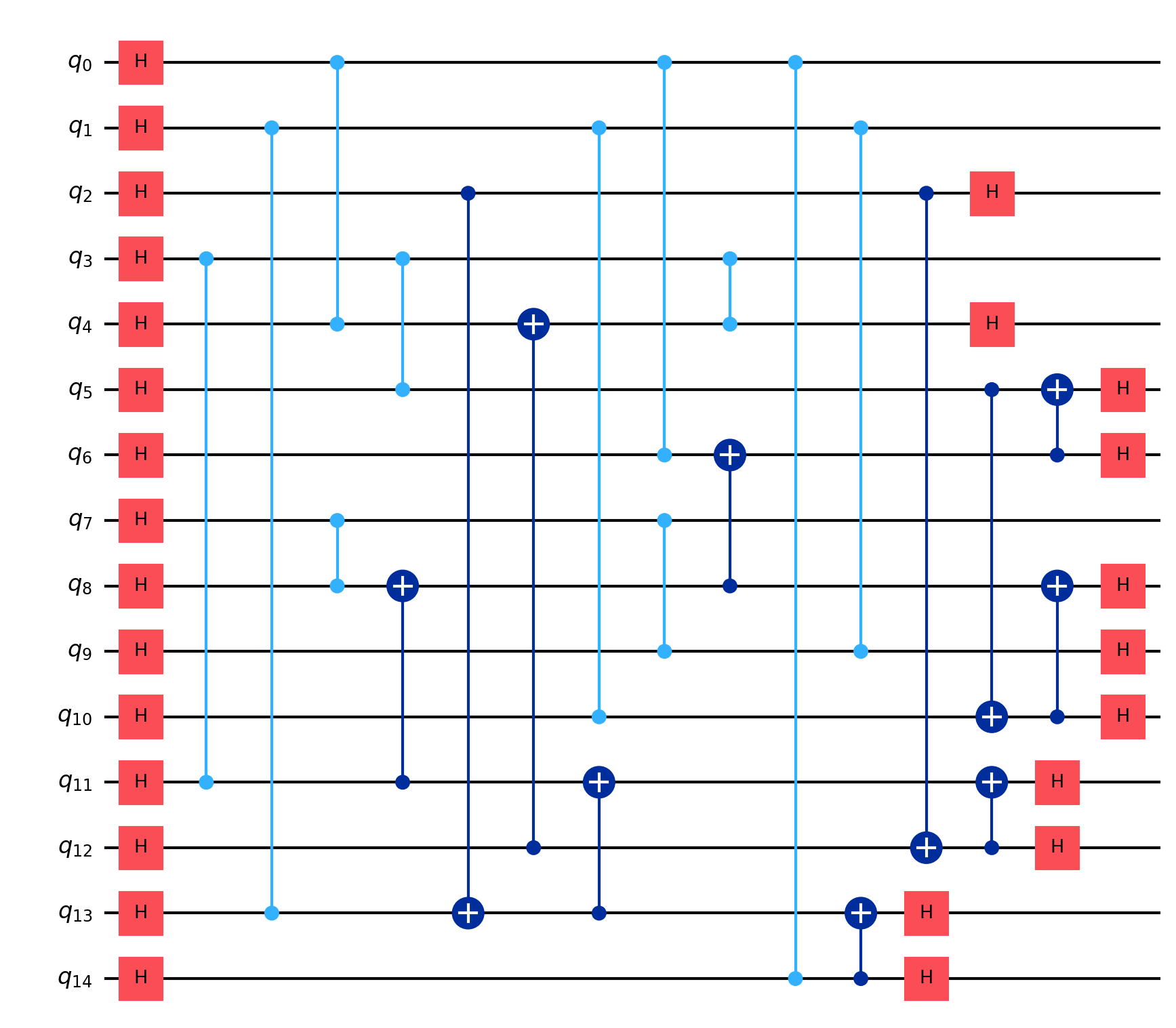}
\caption{Quantum circuit for preparing the $\ket{0}_L$ state of the $\nkd{15}{7}{3}$ Hamming code, obtained with BeFS. The red single-qubit gates are Hadamard; the light blue and dark blue two-qubit gates are the $\CZ$ and $\CX$ entangling gates, respectively.}
\label{fig:}
\end{figure}

\begin{figure}[h!]
\centering
\includegraphics[width=0.99\linewidth]{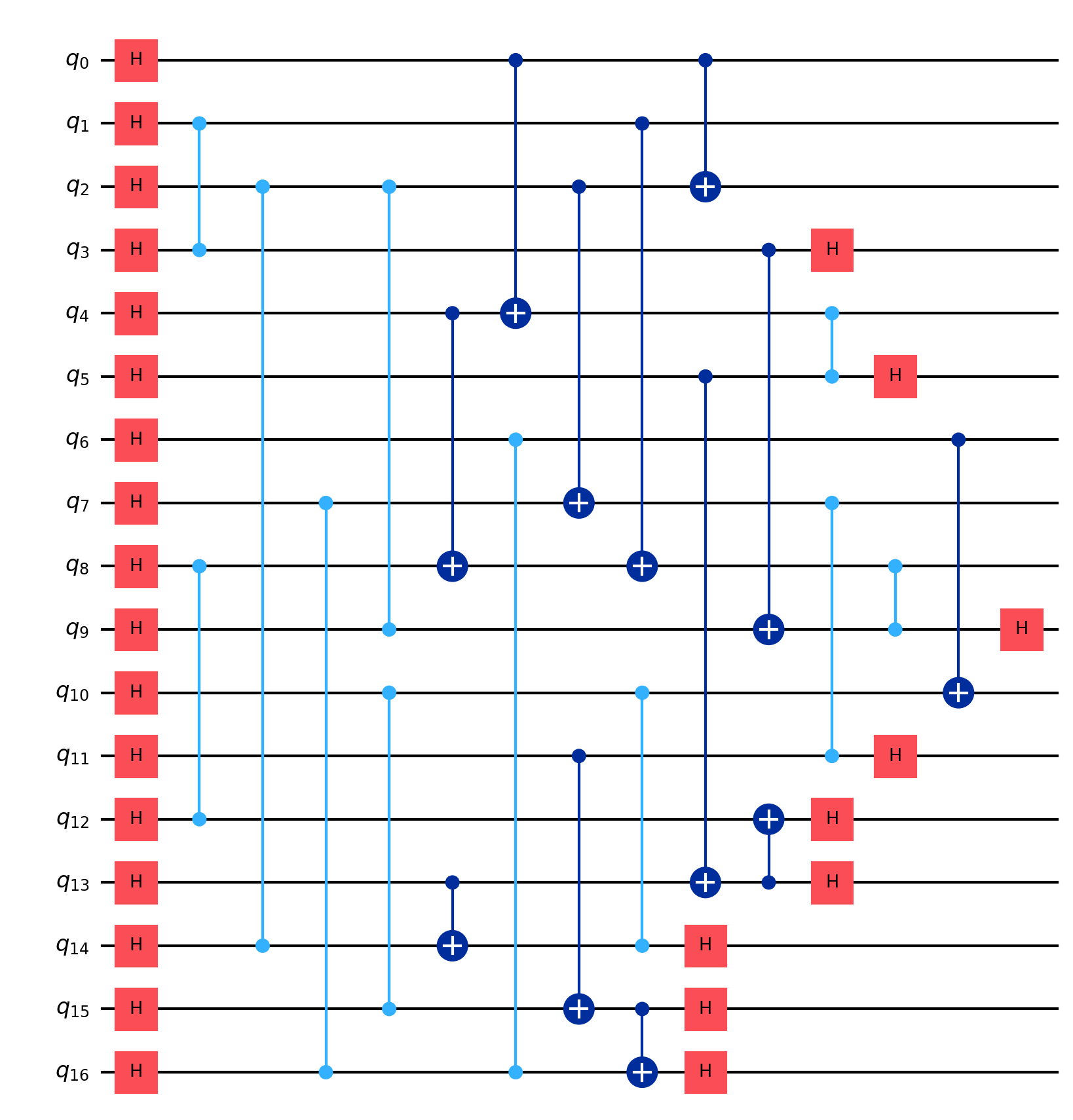}
\caption{Circuit for preparing the $\ket{0}_L$ state of the $\nkd{17}{1}{5}$ color code,  obtained with BeFS. The red single-qubit gates are Hadamard; the light blue and dark blue two-qubit gates are the $\CZ$ and $\CX$ entangling gates, respectively.}
\label{fig:}
\end{figure}

\begin{figure}[h!]
\centering
\includegraphics[width=0.99\linewidth]{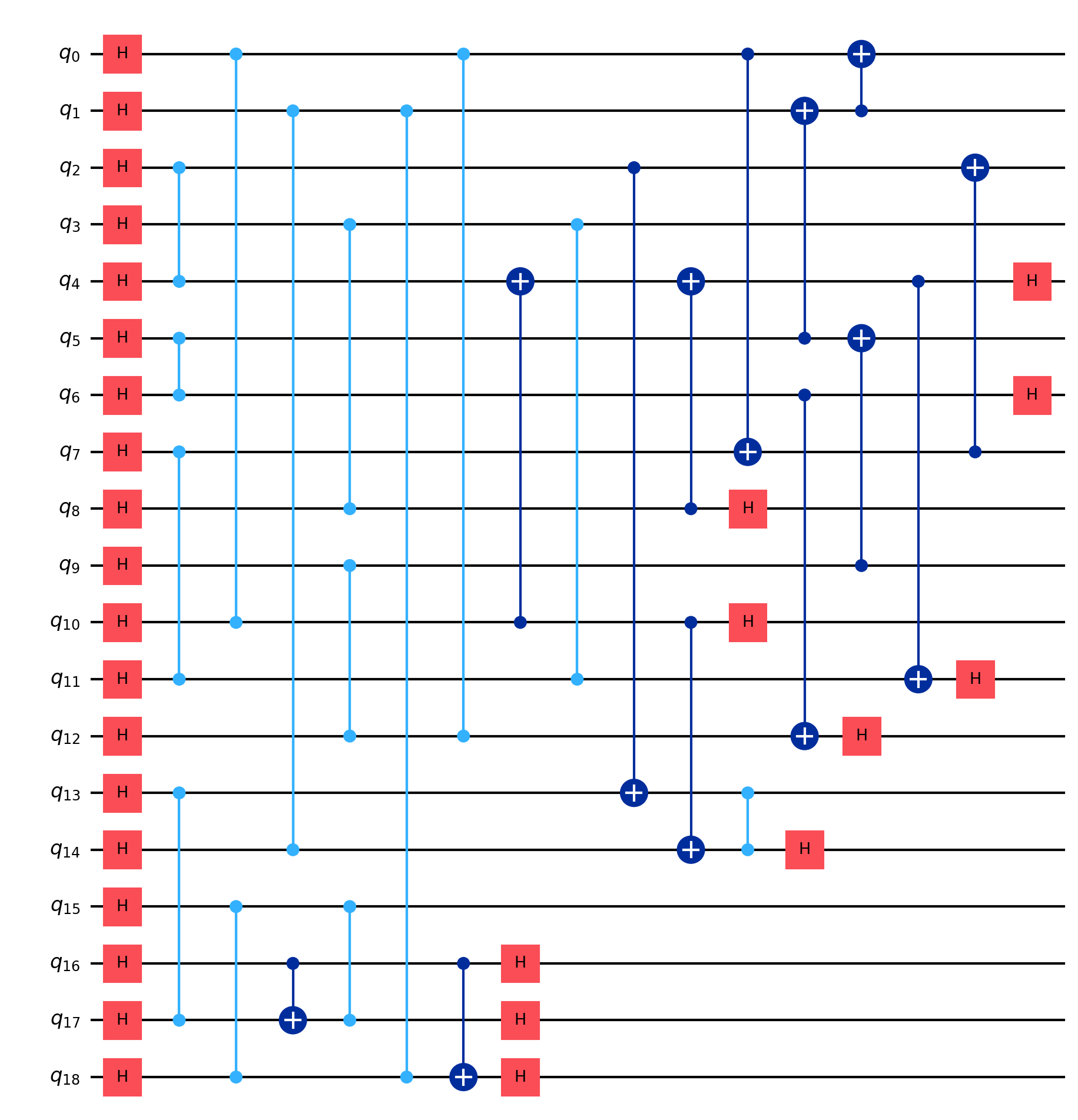}
\caption{Circuit for preparing the $\ket{0}_L$ state of the $\nkd{19}{1}{5}$ color code, obtained with beam search. The red single-qubit gates are Hadamard; the light blue and dark blue two-qubit gates are the $\CZ$ and $\CX$ entangling gates, respectively.}
\end{figure}

\begin{figure*}[htbp]
\centering
\includegraphics[width=17cm]{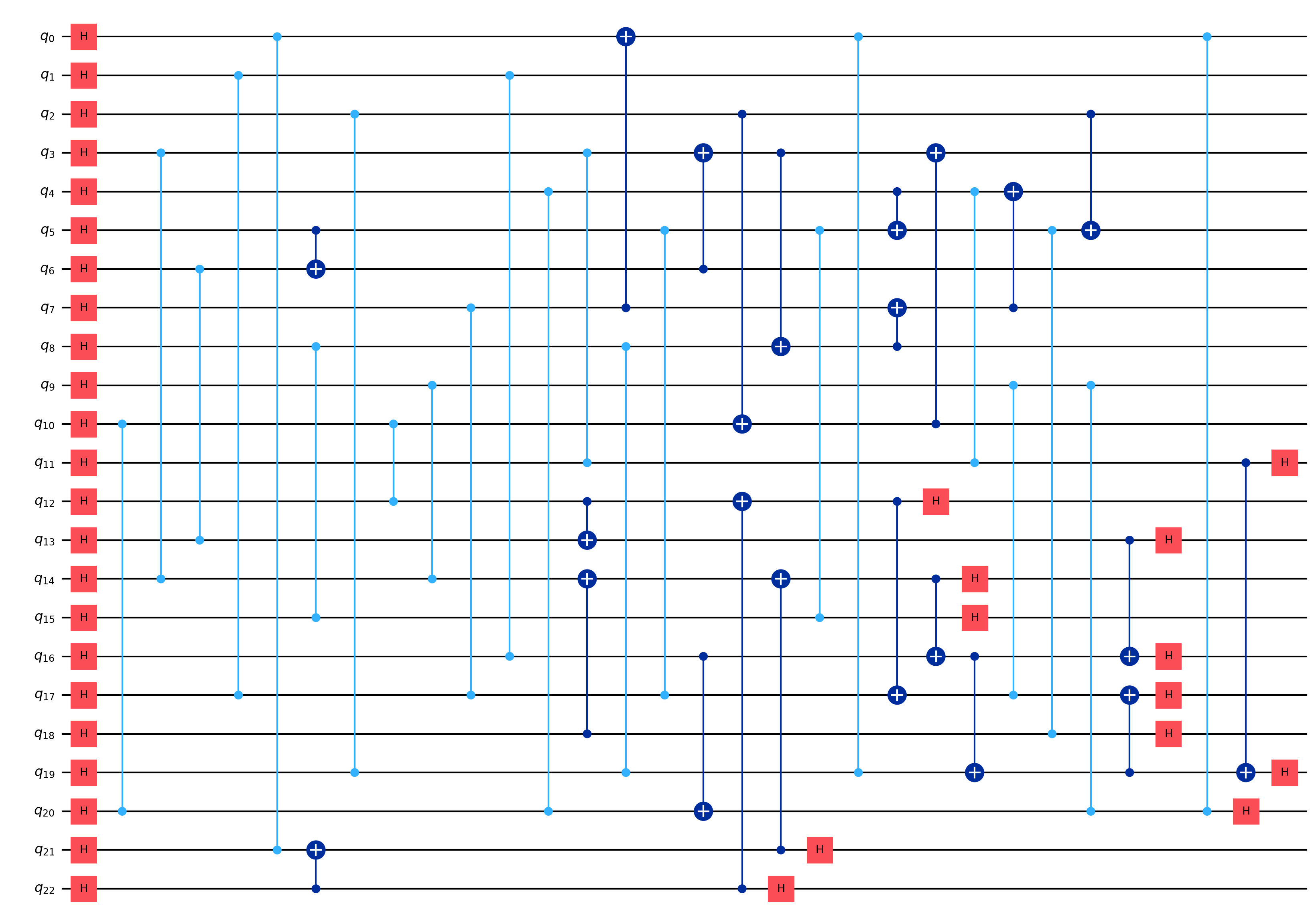}
\caption{Quantum circuit for the preparation of the zero logical state of the $\nkd{23}{1}{7}$ code, with 44 two-qubit gates and two-qubit depth 8, obtained with \textit{QuSynth}. The red single-qubit gates are Hadamard; the light blue and dark blue two-qubit gates are the $\CZ$ and $\CX$ entangling gates, respectively.}
\label{fig:circuit_golay}
\end{figure*}

\begin{figure*}
\centering
\rotatebox{90}{\includegraphics[width=0.95\textheight]{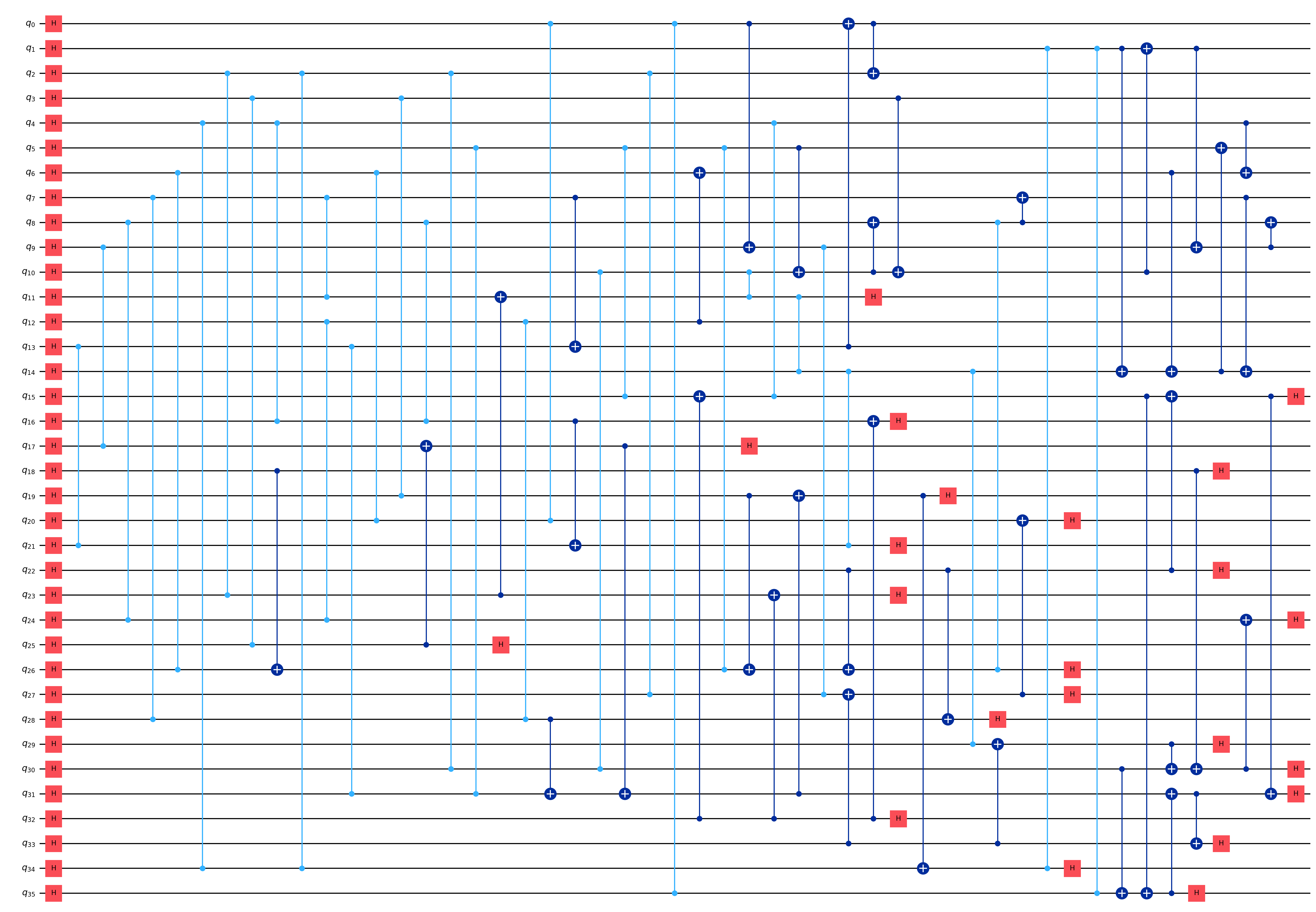}}
\caption{Quantum circuit for the preparation of the zero logical state of the $\nkd{36}{8}{6}$ code, with 77 two-qubit gates and total circuit depth 11, obtained with \textit{QuSynth}. The red single-qubit gates are Hadamard; the light blue and dark blue two-qubit gates are the $\CZ$ and $\CX$ entangling gates, respectively.}
\label{fig:circuit_stab36}
\end{figure*}

\end{document}